\documentclass[12pt]{article} 
 \pdfoutput=1

\usepackage{amsmath,amssymb,amsfonts}
\usepackage{psfrag}
\usepackage{color}
\definecolor{darkblue}{rgb}{0.1,0.1,.7}
\usepackage[colorlinks, linkcolor=darkblue, citecolor=darkblue, urlcolor=darkblue, linktocpage]{hyperref} 
\usepackage[square, comma, sort&compress,numbers]{natbib}
\usepackage[]{amsmath}
\usepackage[]{graphicx}
\usepackage[]{latexsym}
\usepackage{geometry}
\usepackage{amscd}
\usepackage{braket}
\usepackage[all,cmtip]{xy}
\usepackage{mathrsfs}
\usepackage[margin=10pt,font=small,labelfont=bf]{caption}
\usepackage{simplewick}
\usepackage{bbold}
\usepackage{changepage}
\usepackage{dsfont}
\usepackage{bm}
\usepackage{float}
\usepackage[utf8]{inputenc}
\usepackage{setspace}

\numberwithin{equation}{section}

\newcommand{\be}{\begin{eqnarray}}
\newcommand{\ee}{\end{eqnarray}}
\newcommand{\bea}{\begin{eqnarray}}
\newcommand{\eea}{\end{eqnarray}}

\newcommand   \SU    {\mathrm{SU}}
\newcommand \U {\mathrm{U}}
\newcommand   \SO    {\mathrm{SO}}

\renewcommand \Re    {\mathop{\mathrm{Re}}}
\renewcommand \Im    {\mathop{\mathrm{Im}}}

\newcommand   \f  {\phi}

\newcommand   \s  {\sigma}

\renewcommand \th {\theta}

\def\beq{\begin{equation}} 
\def\eeq{\end{equation}} 
\def\<{\langle}
\def\>{\rangle}

\begin{document}

\vspace*{-.6in} \thispagestyle{empty}
\begin{flushright}
\end{flushright}
\vspace{.2in} {\Large
\begin{center}
{\bf A Proof of the Conformal Collider Bounds\vspace{.1in}}
\end{center}
}
\vspace{.2in}
\begin{center}
{\bf 
Diego M. Hofman$^{a}$, Daliang Li$^{b}$, David Meltzer$^c$,}\\
{\bf David Poland$^{c,d}$, Fernando Rejon-Barrera$^a$} 
\\
\vspace{.2in} 
$^a$ {\it Institute for Theoretical Physics, University of Amsterdam, Science Park 904,} \\
{\it Postbus 94485, 1090 GL, Amsterdam, The Netherlands}\\
$^b$ {\it  Department of Physics and Astronomy, Johns Hopkins University, Baltimore, MD 21218}\\
$^c$ {\it Department of Physics, Yale University, New Haven, CT 06511}\\
$^d$ {\it School of Natural Sciences, Institute for Advanced Study, Princeton, NJ 08540}
\end{center}

\vspace{.2in}

\begin{abstract}
In this paper, we prove that the ``conformal collider bounds" originally proposed in~\cite{Hofman:2008ar} hold for any unitary parity-preserving conformal field theory (CFT) with a unique stress tensor in dimensions $d\ge 3$. In particular this implies that the ratio of central charges for a unitary 4d CFT lies in the interval $\frac{31}{18} \geq \frac{a}{c} \geq \frac{1}{3}$. For superconformal theories this is further reduced to $\frac{3}{2} \geq \frac{a}{c} \geq \frac{1}{2}$. The proof relies only on CFT first principles - in particular, bootstrap methods -  and thus constitutes the first complete field theory proof of these bounds. We further elaborate on similar bounds for non-conserved currents and relate them to results obtained recently from deep inelastic scattering.
\end{abstract}

\newpage

\setcounter{tocdepth}{2}
\begin{spacing}{.2}
\tableofcontents
\end{spacing}

\newpage

\section{Introduction}
\label{sec:intro}

It is an extremely important open problem to understand the full set of constraints that ultraviolet consistency places on the infrared behavior of quantum field theories. Conformal field theories provide an exciting arena for probing this question, in part due to the existence of powerful additional symmetry, the availability of experimental data in condensed matter and statistical systems, and deep connections with quantum gravity through holography. The recent striking success of the conformal bootstrap in isolating and solving the low-lying spectrum of the 3d Ising model~\cite{ElShowk:2012ht,El-Showk:2014dwa,Kos:2014bka,Simmons-Duffin:2015qma} gives us reason to think that such constraints may be even more restrictive in $d>2$ than previously thought.

One powerful example of such constraints was given in~\cite{Hofman:2008ar}, where it was argued that the coefficients appearing in the three-point functions between currents and stress tensors, $\<JJT\>$ and $\<TTT\>$, are bounded to a finite region of parameter space. These constraints are particularly striking in superconformal field theories where these coefficients are often calculable. The basic argument in~\cite{Hofman:2008ar} behind these bounds is that for any initial state the energy flux measured at infinity integrated over time should be positive. In other words, calorimeters in a ``conformal collider experiment" should pick up positive energies. Notice that the positivity of the spectrum of energy flux operators was not proved in \cite{Hofman:2008ar}, but was only postulated.

These ``conformal collider bounds" were originally conjectured in 4d, where even in non-supersymmetric theories they place highly nontrivial bounds on the central charges $a$ and $c$ appearing in the trace anomaly, while their generalizations to other dimensions was given in~\cite{Buchel:2009sk,Chowdhury:2012km}. For example in 4d, for any CFT (with or without supersymmetry) the central charges must lie in the region:
\be
\frac{31}{18} \geq \frac{a}{c} \geq \frac{1}{3}.
\ee
Since~\cite{Hofman:2008ar}, it has been an open question whether these constraints are consequences of CFT first principles, such as unitarity, associativity of the operator algebra, and causality. Moreover in~\cite{Farnsworth:2015hum} it was recently questioned whether there may exist consistent theories that violate these constraints but satisfy a weaker set of conditions.

While a full proof of these bounds has not been available, significant progress on this issue has been made on a number of different fronts. In~\cite{Buchel:2009tt,Hofman:2009ug} it was argued that such constraints follow from causality in holographic large $N$ theories by using and extending preliminary results in \cite{Brigante:2007nu,Brigante:2008gz}.\footnote{For a very interesting discussion of the origins and consequences of these constraints in the language of bulk AdS physics, see \cite{Camanho:2014apa}.} Also in \cite{Hofman:2009ug}, a CFT argument was presented by making assumptions on the behavior of lightcone operator product expansions and unitarity bounds on non-local operators. Furthermore,  in~\cite{Kulaxizi:2010jt} some suggestive (but incomplete) arguments were given that the bounds were related to unitarity when the CFT is placed at finite temperature. More recently, the bounds and some generalizations to non-conserved operators have been derived in the context of deep inelastic scattering (DIS) in \cite{Komargodski:2016gci} after making the assumption that the DIS amplitude is bounded by a certain power of the kinematical invariant given by $\lim_{x\rightarrow 0}\mathcal{A}(q^2,x) < x^{-2}$. While all these results are extremely interesting, it is important to stress that there was always some important assumption made that went beyond basic CFT principles.

Another perspective comes from analytical studies of the bootstrap in the lightcone limit, which reveal a direct relation between couplings of low-twist operators and the asymptotic behavior of CFT spectra at large spin~\cite{Fitzpatrick:2012yx,Komargodski:2012ek}, extended further in~\cite{Fitzpatrick:2014vua,Kaviraj:2015cxa,Alday:2015ota,Vos:2014pqa,Alday:2015eya,Kaviraj:2015xsa,Fitzpatrick:2015qma,Li:2015rfa,Alday:2015ewa,Li:2015itl,Dey:2016zbg}. In particular, in~\cite{Li:2015itl} analytic lightcone bootstrap arguments were given for correlation functions containing global symmetry currents and the stress tensor, where a direct connection was found between the 3d conformal collider bounds for $\<JJT\>$ and $\<TTT\>$ and negativity of the anomalous dimensions of large spin double-twist operators. 

Finally, another important development occurred in~\cite{Hartman:2015lfa}, where it was demonstrated that CFT unitarity/reflection positivity implies both causality and sum rules leading to constraints on the signs of products of OPE coefficients. These constraints are closely related to the bound on chaos~\cite{Shenker:2014cwa,Maldacena:2015waa} as discussed in the context of CFT correlators in~\cite{Maldacena:2015iua,Fitzpatrick:2016thx,Perlmutter:2016pkf,Turiaci:2016cvo}. The argument was made for scalar 4-point functions in~\cite{Hartman:2015lfa} and was recently generalized to spinning 4-point functions in~\cite{Hartman:2016dxc}. In the latter work a set of constraints, somewhat weaker than the conformal collider bounds, were derived for the coefficients in $\<JJT\>$ and $\<TTT\>$ for CFTs in general dimensions. The argument in~\cite{Hartman:2016dxc} additionally assumed the absence of scalar operators in the $J \times J$ and $T \times T$ OPEs with dimensions $\frac{d}{2} -1 < \Delta < d-2$.

In the present work we will give a complete proof that the conformal collider bounds must hold in any unitary, parity-preserving conformal field theory with a unique stress energy tensor. We combine the basic argument of~\cite{Hartman:2015lfa} with the refined understanding of positivity conditions obtained from generalizing the lightcone bootstrap arguments of~\cite{Li:2015itl} to general dimensions. We will additionally explain why the bounds hold even in the presence of light scalar operators. Generalizing our argument to non-conserved operators, we reproduce the constraints obtained in the context of deep inelastic scattering in~\cite{Komargodski:2016gci}. Throughout, we assume the CFT preserves parity and has a unique conserved, spin two operator, which is the stress tensor.

Our paper is organized as follows. In section~\ref{sec:overview} we give a brief sketch of our argument. In section~\ref{sec:Hartman} we review the argument of~\cite{Hartman:2015lfa} in the context of operators with spin. In sections~\ref{sec:JJT} and~\ref{sec:TTT} we give our argument for the conformal collider bounds as well as derivations of large spin anomalous dimensions for CFTs in general dimensions. In section~\ref{sec:NonConserved} we describe the generalization to 3-point functions between non-conserved spinning operators. Finally, in section~\ref{sec:discussion} we summarize our results and discuss future directions. Details and extensions of our computations are presented in appendices~\ref{sec:SpinningBlocks}, \ref{sec:NonzeroN}, and \ref{sec:TensorBasis}.

\section{Overview}
\label{sec:overview}

In this section, we start by providing a sketch of the derivation of the conformal collider bounds on the coefficients in $\<J^\mu J^\nu T^{\rho\sigma}\>$ in a unitary conformal field theory. We will make our argument for CFTs with a unique stress tensor and no conserved currents with spin $\ell >2$.\footnote{It was shown in \cite{Maldacena:2011jn,Alba:2015upa} that the presence of higher spin symmetry forces $\<J^\mu J^\nu T_{\rho\sigma}\>$ to saturate the conformal collider bounds for unitary CFTs in $d\ge3$.} 

In \cite{Hartman:2015lfa}, it was demonstrated that very general bounds on the OPE coefficients in unitary CFTs can be derived from the analyticity of the 4-point function $\<\phi(0) O(z,\bar{z}) O(1) \phi(\infty)\>$. This 4-point function can be expanded in different OPE channels. In the lightcone limit of $\bar{z} \rightarrow 1$, crossing symmetry implies:\footnote{Some prefactors and overall coefficients are omitted in this section. They can be found in later sections.}  
\be
G(z,\bar{z}) = \<\phi(0) O(z,\bar{z}) O(1) \phi(\infty)\> \sim 1 + \lambda_{OOT} \lambda_{\phi\phi T} g_{T}(1-z,1-\bar{z}) +\dots \sim \sum_{h,\bar{h}} a_{h,\bar{h}} z^h \bar{z}^{\bar{h}}, \hspace{0.5cm} 
\ee
where we show the contribution from the identity operator and the stress tensor $T$ in the t-channel. Using analyticity, the t-channel coefficients $\lambda_{OOT} \lambda_{\phi\phi T}$ can be related to an integral of $\text{Re}(G(z,\bar{z}) - G(z e^{-2\pi i},\bar{z}))$ over a domain where $z,\bar{z}$ are real and positive. This integrand is positive because reflection positivity implies $a_{h,\bar{h}}\ge 0$. This leads to a bound on the t-channel OPE coefficients: $\lambda_{OOT} \lambda_{\phi\phi T} \ge 0$. 

We apply the same argument now to the 4-point function $G^{\mu\nu}_J\equiv\<J^\mu \phi\phi J^\nu\>$ and derive bounds on the coefficients of $\<J^\mu J^\nu T^{\rho\sigma}\>$ which appear in the t-channel of this 4-point function. There are two independent coefficients $C_J$ and $\lambda_{JJT}$, where $C_J$ is the current central charge. We move the 4 points to a common 2d plane with conformal transformations and use $\{+,-,t\}$ to represent the two lightcone directions on this plane and an arbitrary perpendicular direction. ($x^{\pm}$ are related to the usual Cartesian coordinates of Euclidean space by $x^{\pm}=x^{1}\pm ix^{0}$.)

We focus first on $\<J^{+}\f\f J^{+}\>$. In the t-channel, the dominant contributions in the lightcone limit come from the identity operator and the $T_{--}$ component of the stress tensor. In the $J^+ \times \f$ channels, only symmetric traceless tensors (STTs) contribute at leading order:
\be
G^{++}_J(z,\bar{z}) \sim -2 + \lambda_{J^+ J^+ T_{--}} \lambda_{\phi\phi T} g^{++}_{T}(1-z,1-\bar{z}) + \ldots \sim G^{++}_{J,STT}(z,\bar{z}).
\ee
Again, one can show that the correlation function is analytic in the region of interest and the power series coefficients of $G^{++}_{J}(z,\bar{z})$ in the $z,\bar{z}$ expansion are negative, $a^{++}_{h,\bar{h}}\le 0$. This implies a bound on the product of t-channel coefficients $\lambda_{J^+ J^+ T_{--}} \lambda_{\phi\phi T} \le 0$, where $\lambda_{\phi\phi T} \le 0$ due to the Ward identity. Expressing $\lambda_{J^+ J^+ T_{--}}$ as a function of $C_J$ and $\lambda_{JJT}$, this constraint becomes
\bea
\label{eq:FreeFermionBoundInSummary}
\lambda_{JJT}\leq \frac{\Gamma (\frac{d}{2}+1)}{2\pi^{\frac{d}{2}}} C_J .
\eea
This is the upper half of the conformal collider bounds on $\langle JJT\rangle$ and is saturated in a theory of free fermions. 

Next, we focus on $\<J^{t}\f\f J^{t}\>$. We will show that in the $J^t\times\f$ channel, two families of operators contribute in the leading lightcone limit, the STTs and the mixed symmetry tensors denoted as A, which have a pair of antisymmetrized indices while the rest of the indices are symmetrized:
\be
\label{eq:CrossingJJffInSummary}
G^{tt}_J(z,\bar{z})\sim 1 + \lambda_{J^t J^t T_{--}} \lambda_{\phi\phi T} g^{tt}_{T}(1-z,1-\bar{z}) + \ldots \sim G^{tt}_{J,STT}(z,\bar{z}) + G^{tt}_{J,A}(z,\bar{z}). 
\ee
The same argument will apply, except here the power series coefficients of $G^{tt}_J$ are positive, $a^{tt}_{h,\bar{h}}\ge 0$. This implies  $\lambda_{J^t J^t T_{--}} \lambda_{\phi\phi T} \ge 0$ or $\lambda_{J^t J^t T_{--}}\le0$. However this bound is not optimal, which is expected since we have two analytic functions of $z,\bar{z}$ on the right hand side and each has a power series with positive coefficients. Indeed, we can get a stronger bound by subtracting $G^{tt}_{J,STT}(z,\bar{z})$ from both sides of the crossing equation (\ref{eq:CrossingJJffInSummary}), leading to:
\be
\frac{d-2}{d-1} + \lambda_{J^t J^t T_{--}} \lambda_{\phi\phi T} g^{tt}_{T}(1-z,1-\bar{z}) + \frac{1}{d} \lambda_{J^+ J^+ T_{--}}\lambda_{\phi\phi T}  g^{++}_{T}(1-z,1-\bar{z}) \sim G^{tt}_{J,A}(z,\bar{z}).\hspace{0.5cm} 
\ee
On the left hand side, we have used conformal symmetry to relate $G^{tt}_{J,STT}(z,\bar{z})$ to $G^{++}_{J,STT}(z,\bar{z})$, which is in turn written in terms of the t-channel conformal blocks using crossing symmetry. Analyticity and reflection positivity of $G^{tt}_{J,A}(z,\bar{z})$ then imply:
\bea
\lambda_{JJT}\ge\frac{(d-2) \Gamma (\frac{d}{2}+1)}{2(d-1) \pi ^{\frac{d}{2}} } C_{J}.
\eea
This bound is saturated in a theory of free bosons. Together with (\ref{eq:FreeFermionBoundInSummary}), we have obtained the conformal collider bounds on $\<JJT\>$ that imply energy flux positivity.  

The derivation of the conformal collider constraints for $\<TTT\>$ from $\<T\f\f T\>$ exactly mirrors that of $\<J\f\f J\>$. The main difference is that for $d\geq4$, $\<TTT\>$ depends on three parameters and we will need to consider three crossing relations, corresponding to $\<T^{++}\f\f T^{++}\>$, $\<T^{+3}\f\f T^{+3}\>$, and $\<(T^{33}-T^{44})\f\f(T^{33}-T^{44})\>$. Due to extra degeneracies, $\<TTT\>$ in $d=3$ depends on two parameters (assuming parity), so we will only need to consider the first two equations. In both cases we will re-derive the full conformal collider bounds. 

In the remainder of the paper we will give the above argument in more detail.

\section{Scalar Correlators}
\label{sec:Hartman}

In this section we review the constraints obtained by analyzing the 4-point function of scalar operators, following \cite{Hartman:2015lfa}:
\bea
\label{eq:4PF}
G(z,\bar{z})=\<\f(0)O(z,\bar{z})O(1)\f(\infty)\>.
\eea
In the Euclidean region where $\bar{z}=z^*$, this 4-point function may only have singularities as $z\rightarrow 0, 1, \infty$, as pairs of operators approach each other. They are generically branch points that make the 4-point function multi-valued for independent complex $z$ and $\bar{z}$. 
The first sheet of $G(z,\bar{z})$ embeds the Euclidean region. The second sheet, $G(z e^{-2\pi i},\bar{z})$, is obtained from the first sheet by taking $z$ around the branch point at $0$. 

In this work, we will use the properties of this 4-point function in a very small region near $(z,\bar{z})\sim (1,1)$. Following the notation of~\cite{Hartman:2015lfa}, let us define
\begin{align}
z &= 1 + \sigma,\\
\bar{z} &= 1 + \eta \sigma,
\end{align}
where $\sigma$ is complex with $\Im(\sigma)\ge 0$ and $|\sigma| \le R$, while $\eta$ is real and satisfies $0<\eta\ll R\ll 1$. On the $\sigma$ plane, this is a small half disc above $\sigma=0$ and we exclude the origin from it. We refer to this as region $D$. 
\begin{figure}
\centering
\includegraphics[width=120mm]{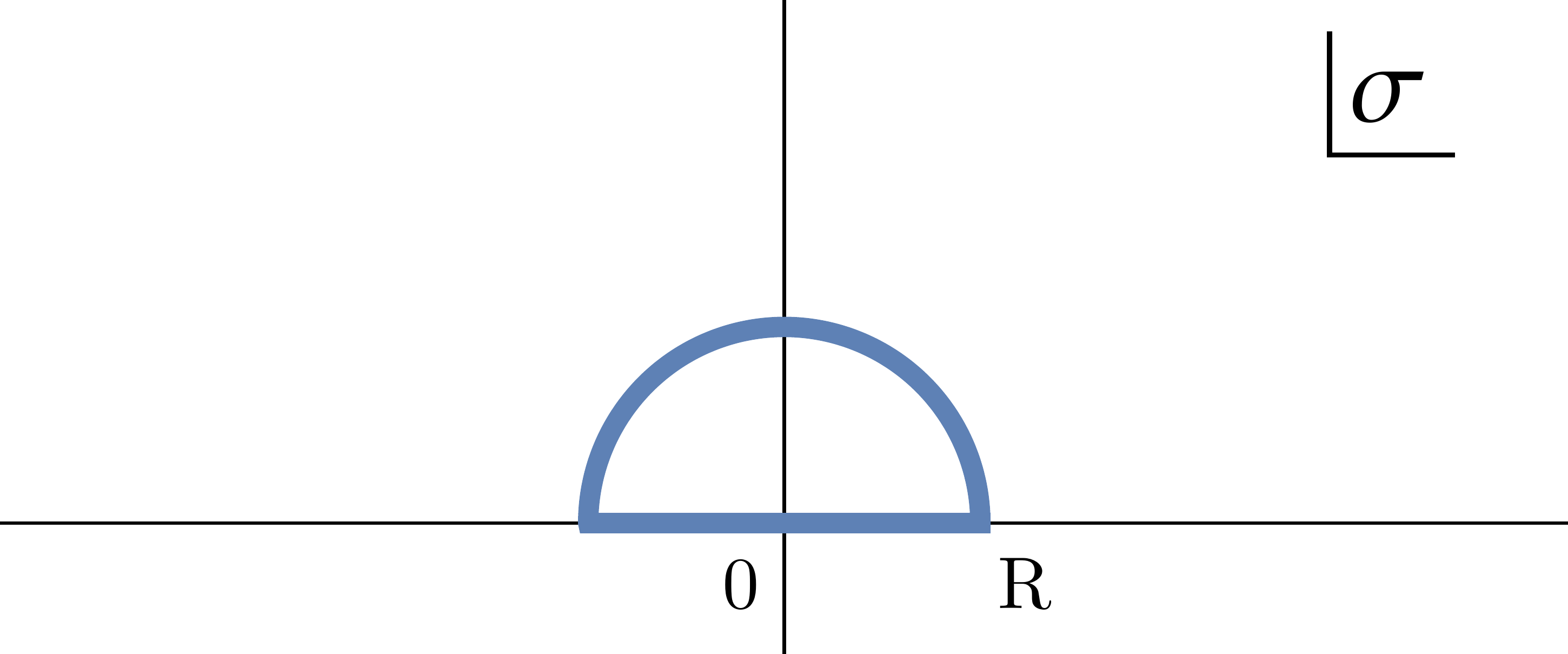}
\caption{\small The region $D$ is a small half disc on the complex $\sigma$ plane above the origin (which is excluded). Here $0<\eta\ll R\ll 1. $\label{fig:D}}
\end{figure}
We define the normalized 4-point functions on the first and the second sheet: 
\begin{align}
G_{\eta}(\s)&\equiv \frac{\<\f(0)O(z,\bar{z})O(1)\f(\infty)\>}{\<\f(0)\f(\infty)\>\<O(z,\bar{z})O(1)\>}=(\eta\s^{2})^{\Delta_{O}}G(1+\s,1+\eta\s),
\\
\widehat{G}_{\eta}(\s)&\equiv \frac{\<\f(0)O(z e^{-2\pi i},\bar{z})O(1)\f(\infty)\>}{\<\f(0)\f(\infty)\>\<O(z,\bar{z})O(1)\>}=(\eta\s^{2})^{\Delta_{O}}G((1+\s)e^{-2\pi i},1+\eta\s).
\end{align}
We will show that both $G_\eta(\sigma)$ and $\widehat{G}_\eta(\sigma)$ are analytic in $D$ and finite at $\sigma=0$. 

On the first sheet, we can expand $G(z,\bar{z})$ in three different OPE channels: 
\begin{align}
\text{s-channel: \qquad} &G(z,\bar{z})=(z\bar{z})^{-\frac{1}{2}(\Delta_{O}+\Delta_{\f})}\sum_{\mathcal{O}}\lambda_{\f O\mathcal{O}}\lambda_{O\f\mathcal{O}}g^{\Delta_{\f O},-\Delta_{\f O}}_{\Delta_{\mathcal{O}},\ell_{\mathcal{O}}}(z,\bar{z}),
\\
\text{t-channel: \qquad} &G(z,\bar{z})=[(1-z)(1-\bar{z})]^{-\Delta_{O}}\sum_{\mathcal{O}}\lambda_{O O\mathcal{O}}\lambda_{\f\f\mathcal{O}}g^{0,0}_{\Delta_{\mathcal{O}},\ell_{\mathcal{O}}}(1-z,1-\bar{z}),
\\
\text{u-channel: \qquad} &G(z,\bar{z})=(z\bar{z})^{\frac{1}{2}(\Delta_{\f}+\Delta_{O})}\sum_{\mathcal{O}}\lambda_{\f O\mathcal{O}}\lambda_{O\f\mathcal{O}}g^{\Delta_{\f O},-\Delta_{\f O}}_{\Delta_{\mathcal{O}},\ell_{\mathcal{O}}}(1/z,1/\bar{z}).
\end{align}
The positions of the singularities indicate that these expansions converge for $|z|<1$, $|1-z|<1$, and $|z|>1$ respectively. The analyticity of $G_\eta(\sigma)$ in $D$ is a direct consequence of the convergence of the t-channel OPE expansion. In terms of $\sigma$ and $\eta$, we have
\be
\label{eq:GNear1}
G_{\eta}(\sigma)= 1+ \sum_{\Delta, \ell} a_{\Delta,\ell} \eta^{\frac{1}{2}(\Delta-\ell)} \sigma^\Delta, 
\ee
where the sum runs over all operators appearing in the t-channel expansion, including descendants. It is also obvious from this convergent expansion that $G_{\eta}(0)=1$ is finite since unitarity constrains $\Delta$ to be positive.

Convergence of the t-channel expansion is not enough to guarantee analyticity on the second sheet. This is because taking $z\rightarrow z e^{-2\pi i}$ can have a nontrivial effect on the t-channel sum. For example, we can consider the sum in the lightcone limit with $\eta\rightarrow 0$ and $\sigma$ finite. The t-channel expansion becomes
\be
\label{eq:TOPELightcone}
G(z,\bar{z})=[(1-z)(1-\bar{z})]^{-\Delta_{O}} \left(1+\sum_{\mathcal{O}_m} \lambda_m(1-\bar{z})^{\frac{1}{2}(\Delta_m-\ell_m)} \tilde{g}_{\Delta_m,\ell_m}(1-z) + \ldots \right),\hspace{0.5cm}
\ee
where $\lambda_m=\lambda_{OO\mathcal{O}_m}\lambda_{\f\f\mathcal{O}_m}$ denotes the coefficient of the contributions of minimal twist operators $\mathcal{O}_m$. We have omitted higher order terms in $\eta$. The lightcone conformal block $\tilde{g}_{\Delta_m,\ell_m}(1-z)$ is regular on the first sheet around $z=1$, but develops singularities on the second sheet. For example, the lightcone conformal block for the stress tensor in 4-dimensions is given by
\be
\tilde{g}_{4,2}(1-z) = -\frac{15 \left[3(1-z^2)+(1+4z+z^2)\log(z)\right]}{2(1-z)^2}.
\ee
It is easy to see that this function is regular as $z\rightarrow 1$, but a pole emerges from the $\log$ term as $z\rightarrow z e^{-2\pi i}$. 

More generally, using the t-channel conformal blocks, $\widehat{G}_{\eta}(\s)$ in the limit $\eta\ll |\s|\ll1$ takes the form
\bea
\label{eq:GhatOPELightcone}
\widehat{G}_{\eta}(\s) = 1-i \hat{\lambda}_m \frac{\eta^{\frac{1}{2}(\Delta_m-\ell_m)}}{\s^{\ell_m-1}}+\ldots,
\eea
where $\mathcal{O}_m$ is the minimal twist operator of largest spin and 
\be
\hat{\lambda}_{m} = \lambda_m \times \frac{2^{1-\ell_m} \pi \Gamma(\Delta_m + \ell_m)^2}{(\Delta_m + \ell_m -1)\Gamma(\frac12 (\Delta_m+\ell_m))^4}.
\ee
The appearance of singularities in the conformal blocks indicate that individual terms in the t-channel sum may increase as we move to the second sheet. So convergence on the first sheet does not imply convergence on the second sheet. These singularities, however, do not indicate that $\widehat{G}_{\eta}(0)$ is divergent. In fact, when $\sigma < \eta$, similar contributions from the omitted terms in (\ref{eq:TOPELightcone}) and (\ref{eq:GhatOPELightcone}) will come in and cancel the divergence. 

We will show that $\widehat{G}_{\eta}(\sigma)$ is analytic in $D$ and $\widehat{G}_{\eta}(0)$ is finite using positivity and convergence of the s- and u-channel expansions. The 4-point function is reflection positive in the s-channel and this implies that the expansion of the 4-point function around $(z,\bar{z})\sim(0,0)$ has positive coefficients: 
\bea
G(z,\bar{z})=(z\bar{z})^{-\frac{1}{2}(\Delta_{O}+\Delta_{\f})}\sum_{h,\bar{h}>0}a_{h,\bar{h}}z^{h}\bar{z}^{\bar{h}},\hspace{1cm} a_{h,\bar{h}}\geq0.
\eea
This was shown in \cite{Hartman:2015lfa} by considering the following state in a radially quantized Hilbert space: 
\bea
\ket{f}\equiv\int_{0}^{1} dr_{1}\int_{0}^{2\pi}d\theta f(r_{1},\theta_{1})O(r_{1}e^{i\th_{1}},r_{1}e^{-i\th_{1}})\f(0)\ket{0}.
\eea
Reflection positivity in radial quantization states that $\bra{f}f\rangle\geq0$ and requiring this hold for arbitrary $f(r,\theta)$ implies $a_{h,\bar{h}}\ge0$. 

In fact, this positivity condition can be further refined. Following \cite{Fitzpatrick:2012yx}, \cite{Kos:2013tga}, and \cite{SimmonsDuffin:2012uy}, we can insert the projector $|\mathcal{O}| \equiv \sum\limits_{\alpha=\mathcal{O},P\mathcal{O},\ldots}|\alpha\rangle\langle \alpha| $ and still have positivity:
\bea
\langle f|\mathcal{O}|f\rangle \ge 0.
\eea
This implies that each partial wave contribution in the s-channel has positive coefficients in its $z$, $\bar{z}$ expansion:
\bea
\lambda_{O\f\mathcal{O}}\lambda_{\f O \mathcal{O}}g^{\Delta_{\f O},-\Delta_{\f O}}_{\mathcal{O}}(z,\bar{z})=z^{-a}\bar{z}^{-b}\sum_{p,q\in\mathbb{Z}_{+}}b_{p,q}z^{p}\bar{z}^{q}, \qquad b_{p,q}\geq0 ,
\eea
where the powers $a,b$ are related to the scaling dimensions.

In the Euclidean region where $\bar{z}=z^{*}$, the s-channel expansion converges for $|z|<1$. The positivity of $a_{h,\bar{h}}$ immediately implies that for independent complex numbers $z,\bar{z}$ satisfying $|z|, |\bar{z}|<1$, the sum still converges since
\be
\label{eq:GBoundGeneral}
\big|(z\bar{z})^{\frac{1}{2}(\Delta_{O}+\Delta_{\f})} G(z,\bar{z})\big| = \big|\sum_{h,\bar{h}>0}a_{h,\bar{h}}z^{h}\bar{z}^{\bar{h}}\big| \le \sum_{h,\bar{h}>0}a_{h,\bar{h}}|z|^{h}|\bar{z}|^{\bar{h}},\hspace{1cm} |z|, |\bar{z}|<1.
\ee
This implies that $\widehat{G}_{\eta}(\sigma)$ is analytic in the region $D\cap \{|z|, |\bar{z}|<1\}$. 
Restricting to real $z\text{, }\bar{z}\in(0,1)$, we have
\bea
|G(ze^{-2\pi i},\bar{z})|\leq G(z,\bar{z}) \label{eqn:real_line_bound}.
\eea
This in turn yields the inequality
\be
\label{eq:GBoundReal}
\Re (G_\eta(\sigma)- \widehat{G}_\eta(\sigma)) \geq0,\hspace{1.4cm} \sigma \in [-R,0).
\ee
In fact, using radial coordinates~\cite{Hogervorst:2013sma}, the region of convergence for the s-channel expansion can be expanded to the whole complex plane excluding $[1,+\infty)$. As detailed in \cite{Hartman:2015lfa}, the bounds analogous to (\ref{eq:GBoundGeneral}) in the radial coordinate implies that $\widehat{G}_\eta(\sigma)$ is analytic in $D/[0,R]$, while the same argument in the u-channel implies analyticity in $D/[-R,0]$. Combining these two channels, we have shown that $\widehat{G}_\eta(\sigma)$ is analytic in $D$.\footnote{The branch cuts of $G(z,\bar{z})$ with respect to $z$ are chosen in the following way. The first branch cut originates from $z=1$ and lies in the lower half plane. The second branch cut originates from $z=0$ and goes along the negative real axis to connect to the branch point at $z=\infty$. There are no other branch cuts on the $z$ plane. The branch cuts on the $\bar{z}$ plane are chosen in the same way.}

Analyticity of both $G_\eta(\sigma)$ and $\widehat{G}_\eta(\sigma)$ in the region $D$ implies that for $\ell_m \ge 2$ we can write the sum rule
\be
\label{eq:ContourScalar}
\Re \oint_{\partial D} d\sigma \sigma^{\ell_m-2} (\widehat{G}_\eta(\sigma)-G_\eta(\sigma) )=0 .
\ee
This contour is a sum of the half circle $S$ and the real line segment $[-R,R]$. Taking the real part of the integral along the half circle will pick up the residue of the pole in $\sigma$ using the identity
\be
\label{eq:Residual}
\Re i \int_S d\sigma \sigma^n = -\pi \delta_{n,-1}.
\ee
Using (\ref{eq:GNear1}) and (\ref{eq:GhatOPELightcone}), we then have
\be
\label{eq:HalfCircleScalar}
\Re \int_{S} d\sigma \sigma^{\ell_m-2} (\widehat{G}_\eta(\sigma)-G_\eta(\sigma) )= \pi \hat{\lambda}_m \eta^{\frac{1}{2}(\Delta_m-\ell_m)} + O( R^{\ell_m-1} ).
\ee
Together with the sum rule (\ref{eq:ContourScalar}) and the positivity property (\ref{eq:GBoundReal}), this implies that for $\ell_{m}\ge 2$:\footnote{Note that we cannot choose $\ell_m = 0,1$ here, otherwise the integral may not be well defined.}
\be
\label{eq:BoundScalar}
\hat{\lambda}_m = \frac{1}{\pi} \lim_{R\rightarrow 0} \lim_{\eta\rightarrow 0} \eta^{-\frac{1}{2}(\Delta_m-\ell_m)} \int_{-R}^{R} d\sigma \sigma^{\ell_m-2} \Re (G_\eta(\sigma)-\widehat{G}_\eta(\sigma) ) \ge 0,
\ee
where $\hat{\lambda}_m \propto \lambda_{OO\mathcal{O}_m}\lambda_{\f\f\mathcal{O}_m}$. If $\mathcal{O}_m$ is the stress tensor, then we have derived constraints on the OPE coefficients of $\langle \phi\phi T\rangle$ and $\langle OO T\rangle$ using reflection positivity. The condition we get here is trivial and can also be obtained easily using the Ward identity. In  future sections, we will generalize this analysis to 4-point functions involving spinning operators, leading to nontrivial constraints on $\langle JJT \rangle$ and $\langle TTT \rangle$.

Here we would like to emphasize that scalar operators in the t-channel with dimension $\frac{d-2}{2}<\Delta< d-2$ will not change the result. To see this we note that that the discontinuity of the t-channel conformal block along the branch cut $z\in(-\infty,0]$ is purely imaginary. Therefore the integral (\ref{eq:HalfCircleScalar}), due to the Kronecker $\delta$ function in (\ref{eq:Residual}), serves as a projector onto blocks of definite spin $\ell_m$.\footnote{For $d>6$, a finite number of subleading $\eta$ orders in the conformal block for scalar exchange could also dominate over the contribution from the stress tensor. But their $\sigma$ dependence is the same as the leading order. So they cannot be picked up in (\ref{eq:Residual}) either.} This projection is reminiscent of the simplified lightcone OPE structure presented in \cite{Hofman:2009ug}, where the scalar contributions dropped out after integration.

As detailed in \cite{Hartman:2015lfa}, one can show using (\ref{eq:GhatOPELightcone}) that a finite number of conserved higher spin currents in the t-channel will contradict the analyticity of $\widehat{G}_\eta(\sigma)$ in the region $D$. If there are an infinite number of higher spin currents the analytic structure of $\widehat{G}_\eta(\sigma)$ can be changed so (\ref{eq:BoundScalar}) no longer applies. Therefore, among the possible conserved currents, (\ref{eq:BoundScalar}) is only useful for bounding the OPE coefficients involving the stress tensor $T$. 

It has been proven in $d=3$ \cite{Maldacena:2011jn} and in $d\geq 4$ \cite{Alba:2015upa} that in a unitary CFT with a finite central charge and a unique stress tensor satisfying the cluster decomposition principle, the presence of one conserved higher spin current forces the three point functions of the conserved operators to coincide with a free field theory expression. Specializing to our case, this means the conformal collider bounds are saturated for $\<JJT\>$ and $\<TTT\>$. Therefore, although our methods will not apply if the theory contains higher spin currents, with our set of assumptions, the conformal collider bounds also hold for CFTs with a higher spin symmetry.

Finally, one can consider how to derive constraints on the coupling to an exchanged non-conserved higher spin operator. Exactly analogous to how one can isolate the stress tensor block even in the presence of a light scalar, one can use (\ref{eq:HalfCircleScalar}) to isolate the contribution of a higher spin operator and derive constraints on the relevant OPE coefficients. In particular, for each choice of spin, the operator with the smallest twist can be isolated. It would be interesting to compare the constraints from reflection positivity to those derived recently in the context of deep inelastic scattering~\cite{Komargodski:2016gci}, but we postpone this analysis to future work.

\section{Bounds on $\< JJT \>$}
\label{sec:JJT}

The above analysis carries over almost verbatim for the case of spinning operators. Again, the t-channel OPE guarantees that the first sheet correlator $G_\eta(\sigma)$ is analytic in $D$, while the analyticity of $\widehat{G}_\eta (\sigma)$ is ensured by reflection positivity and convergence of the s- and u- channel OPEs. Using the t-channel spinning conformal blocks, we will compute $\widehat{G}_\eta (\sigma)$ for $0<\eta\ll|\sigma|\ll 1$ and the result takes the same functional form as in (\ref{eq:GhatOPELightcone}). We can then use the contour integral of $\widehat{G}_\eta (\sigma)-G_\eta(\sigma)$ along $\partial D$ to relate these coefficients to positive quantities. We will see that this reproduces the conformal collider bounds. 

In this section, we first elucidate the structures of the three conformal block decompositions of $\langle J\phi\phi J\rangle$ and explicitly work out the consequence of reflection positivity in the t- and s-channel. We then demonstrate the extraction of the bounds by applying the contour argument to $\langle J^+\phi\phi J^+\rangle$ and $\langle J^t\phi\phi J^t\rangle$.

\subsection{Crossing Symmetry}
\label{subsec:CrossingJJff}
Let us consider the correlator
\be
G_J^{\mu\nu}(z,\bar{z}) \equiv \<J^\mu(0) O(z,\bar{z}) O(1) J^\nu(\infty)\>.
\ee
This 4-point function can be expanded in s-, t- and u-channel conformal blocks. 
In general, the conformal block expansion for a 4-point function of symmetric traceless fields $O^{(\ell)}(x,\epsilon)=O^{\mu_{1}\ldots\mu_{\ell}}\epsilon_{\mu_{1}}\ldots\epsilon_{\mu_{\ell}}$ with $\epsilon^{2}=0$ takes the form 
\begin{align}
\<O^{\ell_{1}}_{1}(x_{1},\epsilon_{1})&O_{2}^{\ell_{2}}(x_{2},\epsilon_{2})O_{3}^{\ell_{3}}(x_{3},\epsilon_{3})O_{4}^{\ell_{4}}(x_{4},\epsilon_{4})\>=\nonumber
\\
&\frac{1}{x_{12}^{\Delta_{1}+\Delta_{2}}x_{34}^{\Delta_{3}+\Delta_{4}}}\bigg(\frac{x_{24}}{x_{14}}\bigg)^{\Delta_{12}}\bigg(\frac{x_{14}}{x_{13}}\bigg)^{\Delta_{34}}\sum_{\mathcal{O},a,b,p}\lambda^a_{12\mathcal{O}}\lambda^b_{34\mathcal{O}}g_{\mathcal{O},a,b,p}^{\Delta_{12},\Delta_{34}}(z,\bar{z})Q^{p}(\{x_{i},\epsilon_{i}\}),
\end{align}
where $\mathcal{O}$ runs over any operator which can appear in both OPEs, $a,b$ run over different possible 3-point function structures, and $p$ runs over each 4-point function structure $Q^{p}$, which is defined to have weight 0 in all coordinates. 

Specializing to the present case, we have: 
\begin{align}
\text{s-channel: \qquad} &G_J^{\mu\nu}(z,\bar{z})=(z\bar{z})^{-\frac{1}{2}(\Delta_{\f}-\Delta_{J})}\sum_{\mathcal{O}}\lambda_{J\f\mathcal{O}}\lambda_{\f J\mathcal{O}}g^{\Delta_{J\f},\Delta_{\f J},\mu\nu}_{\mathcal{O}}(z,\bar{z}) \label{eq:JJOOsch} \\ \nonumber
&\hspace{1.6cm}= G^{\mu\nu}_{J,STT}(z,\bar{z})+ G^{\mu\nu}_{J,A}(z,\bar{z}),
\\
\text{t-channel: \qquad} &G_J^{\mu\nu}(z,\bar{z})=[(1-z)(1-\bar{z})]^{-\Delta_{\f}}\sum_{\mathcal{O},b}\lambda^b_{JJ\mathcal{O}}\lambda_{\f \f \mathcal{O}}g^{0,0,\mu\nu}_{\mathcal{O},b}(1-z,1-\bar{z})  \label{eq:JJOOtch},
\end{align}
where we have absorbed the tensor structures $Q^p$ into the conformal blocks, since in our configuration they will also become functions of $z$ and $\bar{z}$. This will not be true for generic configurations.
$J$ is a conserved current, so $\Delta_{J}=d-1$. 
In the $J\times\f$ OPE there are in general two families of operators, corresponding to symmetric traceless tensors $STT \equiv \{\mathcal{O}_{[\ell]},\ell\ge 0\}$ and mixed symmetry tensors $A \equiv \{\mathcal{O}_{[\ell,1]}, \ell\ge1\}$, where the latter has a pair of indices antisymmetrized and the other $(\ell - 1)$ indices symmetrized. 

The case $d=3$ is special since all irreducible tensors are equivalent to symmetric traceless tensors through the use of the $\epsilon$ tensor. Since we assume that the theory preserves parity we can still distinguish two families of operators by their parity. In this case we use the label $STT$ for operators that are parity even and $A$ for operators that are parity odd. In both $d=3$ and $d\geq4$, each class of operators has a unique tensor structure in the $J\times \phi$ OPE.

In the $J\times J$ OPE, operators with spin can appear with two independent parity-preserving tensor structures, while scalars have a unique tensor structure. The index $b$ on $\lambda^b_{JJ\mathcal{O}}$ is introduced to account for the generic appearance of multiple 3-point function structures. 

In the lightcone limit where $\bar{z}\rightarrow 1$ and $z$ is finite, the $t$-channel expansion is organized by the twist $\Delta-\ell$. The leading order contribution is given by the identity operator, which has twist zero. 
The identity contribution has a power law singularity $(1-\bar{z})^{-\Delta_{\f}}$ in the lightcone limit, while each single conformal block in the s-channel contains at most a $\log(1-\bar{z})$ singularity. 
Therefore, as established in \cite{Fitzpatrick:2012yx},  leading terms in the t-channel can only be reproduced via infinite sums over the spins of families of operators in the s-channel.
In the present case, there are two such families: $[J\f]^{[\ell]}_{n,\ell} \in STT$, with the schematic form $J^{\mu_1} \partial^{(\mu_2}\dots\partial^{\mu_\ell)}\partial^{2n}\phi$,
and $[J\f]^{[\ell,1]}_{n,\ell} \in A$, with the schematic form $J^{[\mu_1} \partial^{(\mu_2]}\partial^{\mu_3}\dots\partial^{\mu_\ell)}\partial^{2n}\phi$. As $\ell\rightarrow\infty$, the anomalous dimensions of these operators vanish as a power law in $\ell$ and their twists approach $d-2+\Delta_{\f}+2n$ and $d-1+\Delta_{\f}+2n$, respectively.

We will work with two polarizations of the 4-point function, $G^{++}_J$ and $G^{tt}_J$. As demonstrated for $d=3$ in \cite{Li:2015itl}, if we take $z$ to be small then at leading order in $z$ only $[J\f]^{[\ell]}_{0,\ell}$ contributes in the matching of the low-twist t-channel contributions to $G^{++}_J$, while for $G^{tt}_J$ both $[J\f]^{[\ell]}_{n,\ell}$ and $[J\f]^{[\ell,1]}_{n,\ell}$ contribute.\footnote{In \cite{Li:2015itl} the roles of the s- and t-channels are reversed.} Using the spinning conformal blocks, we show in appendix \ref{sec:SpinningBlocks} that this structure holds in general dimensions and persists when $z$ is finite. 

The next-to-leading order contribution in the t-channel comes from the stress tensor conformal block,
which will contain a $\log(z)$ term.\footnote{As discussed in the end of section \ref{sec:Hartman}, the presence of scalars and spin-1 conserved currents in the t-channel does not change the result of our analysis. So to simplify the discussion we ignore their possible contributions. We assume that the CFT we are considering has no higher spin symmetries.}
It is this $\log(z)$ that eventually leads to the $\sigma^{-1}$ enhancement of the correlator on the second sheet when we take $z\rightarrow z e^{-2\pi i}$. 
In the s-channel, this term is reproduced via the anomalous dimensions of the large-spin double-twist operators. The s-channel conformal blocks are proportional to $z^{\frac{\tau}{2}}$, where the twist $\tau$ is given by
\begin{align}
\label{eq:TwistJphiSTT}
\tau_{[J\f]^{[\ell]}_{n,\ell}} &= d-2+\Delta_{\f}+2n + \frac{\gamma_{[J\f]^{[\ell]}_{n,\ell}}}{\ell^{d-2}} + \ldots, \\
\label{eq:TwistJphiA}
\tau_{[J\f]^{[\ell,1]}_{n,\ell}} &= d-1+\Delta_{\f}+2n + \frac{\gamma_{[J\f]^{[\ell,1]}_{n,\ell}}}{\ell^{d-2}} + \ldots.
\end{align}
Expanding $z^{\frac{\tau}{2}}$ to first order in $\frac{1}{\ell^{d-2}}$, we see that a $\log(z)$ term appears. The power $d-2$ ensures that after the large spin sum, this term matches the $1-\bar{z}$ dependence of the $\log(z)$ term in the t-channel stress tensor block. The coefficients $\gamma_{[J\f]^{[\ell]}_{n,\ell}}$ and $\gamma_{[J\f]^{[\ell,1]}_{n,\ell}}$ are also determined by this matching and are functions of the t-channel OPE coefficients appearing in $\langle JJT \rangle$ and $\langle\f\f T\rangle$. 

\subsection{Reflection Positivity}
\label{subsec:ReflectionPositivityJJff}
To elucidate the consequences of reflection positivity, we consider the following states in the Hilbert space of radial quantization
\begin{align}
\ket{f,\epsilon}=&\int_{0}^{1}dr_{1}\int_{0}^{2\pi}d\theta_{1} r_{1}^{\Delta_{J}+\Delta_{\f}}f(r_{1},\theta_{1})\phi(r_{1}e^{i\th_{1}},r_{1}e^{-i\th_{1}})J^{\mu}\epsilon_{\mu}(0)\ket{0},
\\
\bra{f,\epsilon^{*}}=&\bra{0}\epsilon^{*}_{\nu}I^{\nu}_{\rho}(\infty \hat{x}^{1})J^{\rho}(\infty\hat{x}^{1})\int_{0}^{1}dr_{2}\int_{0}^{2\pi}d\theta_{2} r_{2}^{\Delta_{J}-\Delta_{\f}}f^{*}(r_{2},\theta_{2})\phi\bigg(\frac{1}{r_{2}}e^{i\th_{2}},\frac{1}{r_{2}}e^{-i\th_{2}}\bigg).
\end{align}
We used the notation $f(\infty \hat{x}^{1})\equiv \lim\limits_{r\rightarrow\infty} f(r \hat{x}^{1})$ where $\hat{x}^1$ is the unit vector pointing at the direction $x^1 = \frac{1}{2}(x^+ +x^-)$.
Note that the inversion tensor,
\be
I^{\mu\nu}(x) = \eta^{\mu\nu} - 2\frac{x^\mu x^\nu}{x^2},
\ee 
appears in the definition of $\bra{f,\epsilon^{*}}$, and in particular $I^{-}_{+}(\infty\hat{x}^{1}) = -1$, 
which will lead to some additional signs compared to the scalar case. Namely, for $\epsilon_{\mu}$ pointing in the $+$ or the $t$ direction, positivity $\bra{f,\epsilon^{*}} f, \epsilon\rangle \ge 0$ will imply that the power series coefficients of $-G^{++}_J(z,\bar{z})$ or $G^{tt}_J (z,\bar{z})$ will be positive semidefinite when expanded around $(z,\bar{z})\sim(0,0)$. 
Setting $z=1+\sigma$ and $\bar{z}=1+\eta \sigma$, we define the following normalized correlation functions:
\begin{align}
G_{J,\eta}^{\mu\nu}(\s) &\equiv(\eta\s^{2})^{\Delta_{\f}}G_J^{\mu\nu}(1+\s,1+\eta\s),
\\
\widehat{G}_{J,\eta}^{\mu\nu}(\s) &\equiv(\eta\s^{2})^{\Delta_{\f}}G_J^{\mu\nu}((1+\s)e^{-2\pi i},1+\eta\s).
\end{align}
As in the scalar case, these positivity conditions lead to the analyticity of $\widehat{G}_{J,\eta}^{++}(\s)$ and $\widehat{G}_{J,\eta}^{tt}(\s)$ in region $D$ as well as the positivity of $-G_{J,\eta}^{++}(\s) + \widehat{G}_{J,\eta}^{++}(\s)$ and $G_{J,\eta}^{tt}(\s) - \widehat{G}_{J,\eta}^{tt}(\s)$ on the real line segment $\sigma\in[-R,R]$. 

We can sharpen these conditions by inserting the projector $|\mathcal{O}|$ into the norm. Since $\<f, \epsilon^*|\alpha\>\< \alpha|f,\epsilon\> \ge 0$ for any state $\alpha$ in the conformal multiplet of $\mathcal{O}$, we see that each conformal block $\lambda_{J\f \mathcal{O}} \lambda_{\f J \mathcal{O}}G^{\mu\nu}_{\mathcal{O}}$ has negative coefficients in the $z,\bar{z}$ expansion if $\mu=\nu=+$ and positive coefficients if $\mu=\nu=t$. Therefore, the analyticity and boundedness still holds on the second sheet for this partial contribution to the 4-point function. 

Finally, we note that the results in this subsection do not depend on taking the lightcone limit. In particular, they hold when $|\sigma| < \eta$.

\subsection{$\langle J^+ \phi\phi J^+ \rangle$}
\label{subsec:J+J+ff}
Using the t-channel spinning conformal blocks for stress tensor exchange, we can derive the correlation function on the second sheet $\widehat{G}^{++}_{\eta}(\s)$ at next-to-leading order in $\eta$:  
\begin{align}
 &\widehat{G}^{++}_{\eta}(\sigma)=-2C_{J}\notag\\
&\quad+i \lambda_{\f\f T}\frac{d (d-2) \Gamma(d+1) \left[2^{d+2} \pi ^{\frac{d+1}{2}} \Gamma \left(\frac{d+3}{2}\right)\lambda_{JJT} -\pi \Gamma (d+2)C_{J} \right]}{\pi ^{\frac{d}{2}}\sqrt{C_{T}}  \Gamma \left(\frac{d}{2}+1\right)^3} \frac{\eta ^{\frac{d}{2}-1}}{\sigma}+\ldots, \label{eq:hatG++}
\\
&\lambda_{\f\f T} = -\frac{d\Delta_{\f}}{d-1}\frac{1}{\sqrt{C_{T}}}.
\end{align}
Up to the sign flip from the inversion tensor, this has the same form as the scalar case, $\widehat{G}_{\eta}(\s)=1-i\hat{\lambda}\eta^{\frac{\tau}{2}}\s^{-1}$. Similarly, the contour integral implies the sum rule
\begin{align}
\pi \Gamma (d+2)C_{J} -  2^{d+2} \pi ^{\frac{d+1}{2}} \Gamma \left(\frac{d+3}{2}\right)\lambda_{JJT} =& \\ \nonumber
 A \lim_{R\rightarrow 0} \lim_{\eta\rightarrow 0} \eta^{1-\frac{d}{2}} \int_{-R}^{R} d\sigma & \Re(-G_{J,\eta}^{++}(\sigma)+\widehat{G}_{J,\eta}^{++}(\sigma) ) \ge 0,
\end{align}
where $A$ is a positive constant. After some simplification, we arrive at the inequality
\bea\label{eq:JJTbound1}
\lambda_{JJT}\leq\frac{\Gamma \left(\frac{d}{2}+1 \right)}{2\pi^{\frac{d}{2}}} C_{J} .
\eea
This inequality is one of the conformal collider bounds, saturated by a theory of free fermions.

\subsection{$\langle J^t  \phi\phi J^t \rangle$}
\label{subsec:JtJtff}

To find the free boson bound, we need to remove the contribution from all symmetric traceless tensors to $G^{tt}(z,\bar{z})$: 
\begin{align}
\label{eq:GJttSubtracted}
G^{tt}_{J,A}(z,\bar{z}) = G^{tt}_{J}(z,\bar{z})-G^{tt}_{J,STT}(z,\bar{z}).
\end{align}
As mentioned in section \ref{subsec:ReflectionPositivityJJff}, $G^{tt}_{J,A}(z,\bar{z})$ is also analytic on the second sheet and is bounded by its magnitude on the first sheet.

The functions $G^{tt}_{J,STT}(z,\bar{z})$ and $G^{++}_{J,STT}(z,\bar{z})$ are related by conformal symmetry and this relation can be worked out order-by-order in $(1-\bar{z})$ using the explicit form of the spinning conformal blocks. To leading and next-to-leading order in $(1-\bar{z})$, the ratio of the two contributions is independent of $z$. We show in appendix \ref{sec:polratios} that the ratios between them are $\frac{1}{2}\frac{1}{1-d}$ at order $(1-\bar{z})^{-\Delta_{\f}}$ and $-\frac{1}{d}$ at order $(1-\bar{z})^{\frac{d-2}{2}-\Delta_{\f}}$. 
Furthermore, the leading lightcone singularities of $G^{++}_{J}(z,\bar{z})$ from the exchange of the identity and the stress tensor operators come exclusively from the infinite sum over $STT$ double twist operators in the s-channel. Therefore, in the lightcone limit $G^{++}_{J}(z,\bar{z})$ is related in a simple way to $G^{tt}_{J,STT}(z,\bar{z})$. 
The subtraction (\ref{eq:GJttSubtracted}) can then be computed explicitly in the small $\eta$ limit. After continuing to the second sheet we find 
\begin{align}
&\widehat{G}^{tt}_{J,A,\eta}(\sigma)\equiv(\eta\s^{2})^{\Delta_{\f}}G^{tt}_{J,A}((1+\s)e^{-2\pi i},1+\eta\s)=\frac{d-2}{d-1}C_{J}\notag\\
&-i \lambda_{\f\f T}\frac{2 \Gamma (d+2) \left[2^{d+1} \pi ^{\frac{d+1}{2}}(d-1)  \Gamma \left(\frac{d+1}{2}\right)\lambda_{JJT} - \pi (d-2) \Gamma(d+1)C_{J}\right]}{\pi ^{\frac{d}{2}}\sqrt{C_{T}}  \Gamma \left(\frac{d}{2}+1\right)^3} \frac{\eta ^{\frac{d}{2}-1}}{\sigma} +\ldots.
\end{align}
Using a contour integral similar to (\ref{eq:ContourScalar}), reflection positivity then implies the inequality
\bea\label{eq:JJTbound2}
\lambda_{JJT}\ge\frac{(d-2) \Gamma (\frac{d}{2}+1)}{2(d-1)\pi^{\frac{d}{2}}} C_{J}.
\eea
This inequality is the other conformal collider bound on $\langle JJT\rangle$, saturated by a theory of free bosons.

Finally, as in \cite{Hofman:2008ar}, the supersymmetric conformal collider bounds follow from the general bounds derived above. If the current does not correspond to the R symmetry then it is contained in a multiplet with a scalar. In this case supersymmetry fixes $\lambda_{JJT}$ in terms of $C_{J}$ via the relation \cite{Osborn:1993cr,Hofman:2008ar}
\bea 
\lambda_{JJT}= \frac{d(d-2) \Gamma \left(\frac{d}{2}+1\right)}{2 (d-1)^2 \pi ^{\frac{d}{2}}} C_{J},
\eea
which satisfies the conformal collider bounds.

If we consider 4d $\mathcal{N}=1$ SCFTs and $J$ is the superconformal $\U(1)_R$ current, then we instead have\footnote{These relations are straightforwardly derived using the covariant formalism of~\cite{Park:1997bq,Osborn:1998qu} or in superembedding space~\cite{Goldberger:2011yp,Siegel:2012di,Maio:2012tx,Kuzenko:2012tb,Goldberger:2012xb,Khandker:2012pa,Fitzpatrick:2014oza,Khandker:2014mpa,Li:2016chh}.}
\bea
\lambda_{JJT}=\frac{2 (a+3c)}{9c \pi ^2} C_{J} \quad \Rightarrow \quad \frac{3}{2}\geq\frac{a}{c}\geq0 \label{eq:SUSY1ac},
\eea
where $a$ is the Euler anomaly and $c$ is proportional to the central charge $C_{T}$ (see appendix \ref{sect:TTT} for the precise relation). Notice that, in this case, this is not the strongest lower bound in a $\mathcal{N}=1$ SCFT. A stronger bound, $\frac{3}{2} \geq \frac{a}{c} \geq \frac{1}{2}$, comes from looking at the the stress tensor, see section \ref{sec:TTT}.

For 4d $\mathcal{N}=2$ SCFTs, if $J$ is the superconformal $\SU(2)_R$ current, we have instead
\bea
\lambda_{JJT}=\frac{4(a+c)}{9c \pi ^2} C_{J} \quad \Rightarrow \quad \frac{5}{4}\geq \frac{a}{c}\geq \frac{1}{2}.\label{eq:SUSY2ac}
\eea 
These constitute the strongest bounds on the ratio $\frac{a}{c}$ for 4d $\mathcal{N}=2$ theories. In 4d $\mathcal{N}=4$ SCFTs $a=c$, so the bounds are always satisfied.

\subsection{Anomalous Dimensions}
\label{subsec:JfAnomalousDimJ}

We will now show that following the above analysis, the large $\ell$ anomalous dimensions of the $[J\f]^{[\ell]}_{0,\ell}$ and $[J\f]^{[\ell,1]}_{0,\ell}$ double-twist operators due to the exchange of the stress tensor are negative semidefinite, generalizing the $d=3$ results of~\cite{Li:2015itl} to arbitrary dimensions. Matching the t-channel identity contribution in the s-channel yields the large $\ell$ asymptotics of the OPE coefficient for double twist operators:
\begin{align}
(\lambda_{J\f[J\f]^{[\ell]}_{0,\ell}})^{2}&=\frac{C_{J}\sqrt{\pi } 2^{-\Delta_{\f}-d+5}}{2^{\ell}\Gamma (\Delta_{\f}) \Gamma (d)}\ell^{\frac{1}{2} (2 \Delta_{\f}+2d-7)},
\\
(\lambda_{J\f[J\f]^{[\ell,1]}_{0,\ell}})^{2}&=\frac{C_{J}\sqrt{\pi } (d/2-1) 2^{-\Delta_{\f}-d+4}}{2^{\ell}\Gamma (\Delta_{\f}) \Gamma (d)}\ell^{\frac{1}{2} (2 \Delta_{\f}+2d-5)}.
\end{align}
Matching the $\log(z)$ term that comes from the stress tensor contribution in the t-channel, we can find the large $\ell$ anomalous dimensions of double-twist operators in the s-channel. The coefficients in (\ref{eq:TwistJphiSTT}) and (\ref{eq:TwistJphiA}) for $n=0$ are given by
\begin{align}
\gamma_{[J\f]^{[\ell]}_{0,\ell}}&=\lambda_{\f\f T}\frac{(d-2)  \Gamma(d+1) \Gamma (d+2) \Gamma (\Delta_{\f}) \left[ d \Gamma\left(\frac{d}{2}\right)C_{J}-4  \pi ^{\frac{d}{2}}\lambda_{JJT}\right]}{16 \pi^{\frac{d}{2}}\sqrt{C_{T}} C_{J} \Gamma \left(\frac{d}{2}+1\right)^3 \Gamma \left(-\frac{d}{2}+\Delta_{\f}+1\right)}\frac{1}{\ell^{d-2}},
\\
\gamma_{[J\f]^{[\ell,1]}_{0,\ell}}&=\lambda_{\f\f T}\frac{ \Gamma(d+1) \Gamma (d+2) \Gamma (\Delta_{\f}) \left[2  (d-1) \pi ^{\frac{d}{2}}\lambda_{JJT}- (d-2) \Gamma \left(\frac{d}{2}+1\right)C_{J}\right]}{4 \pi^{\frac{d}{2}}\sqrt{C_{T}} C_{J} (d-2) \Gamma \left(\frac{d}{2}+1\right)^3 \Gamma \left(-\frac{d}{2}+\Delta_{\f}+1\right)}\frac{1}{\ell^{d-2}}.
\end{align}
Comparing with the inequalities~(\ref{eq:JJTbound1}) and~(\ref{eq:JJTbound2}), this proves that in the large $\ell$ limit, the symmetric traceless double-twist states $[J\f]^{[\ell]}_{0,\ell}$ and the mixed symmetry double-twist states $[J\f]^{[\ell,1]}_{0,\ell}$ have negative anomalous dimensions arising from the exchange of the stress tensor. 

In fact, extending these formulas to all $\gamma_{[J\f]^{[\ell]}_{n,\ell}}$ and $\gamma_{[J\f]^{[\ell,1]}_{n,\ell}}$ in the regime when $\ell\gg n \geq 0$, the anomalous dimensions are always proportional to the same linear combination of t-channel OPE coefficients. We have explicitly computed these anomalous dimensions in appendix~\ref{sec:NonzeroN}. We find that they are all negative semidefinite because of the conformal collider bounds. In a quantum gravitational theory in AdS dual to a CFT, the double-twist states correspond to two-particle bound states and the anomalous dimensions from $T$ exchange correspond to the gravitational binding energy between the particles. We have therefore proven that due to unitarity and crossing symmetry of the CFT, gravity must be attractive in AdS between a scalar particle and a gauge boson separated at super-horizon distances. Note that this result does not rely on any large $N$ limit.

The universal attractive character of gravity at long distances is precisely violated by Gauss-Bonnet theories of AdS gravity with large higher derivative corrections. It was precisely this fact that led originally to bounds on shear viscosity in \cite{Brigante:2007nu,Brigante:2008gz} which were later generalized and reinterpreted in terms of conformal collider bounds in \cite{Hofman:2008ar,Buchel:2009tt,Hofman:2009ug}. The argument above establishes universal attractiveness of gravity as an inevitable consequence of the holographic principle.

\section{Bounds on $\< TTT \>$}
\label{sec:TTT}

In this section, we apply a similar argument to what was used in section \ref{sec:JJT} to a 4-point function containing two scalars $\phi$ and two stress tensors $T$:
\be
G_T^{\mu\nu\rho\sigma}(z,\bar{z}) \equiv \<T^{\mu\nu}(0) \phi(z,\bar{z}) \phi(1) T^{\rho\sigma}(\infty)\>.
\ee
We will show that the conformal collider bounds on the coefficients in $\langle TTT\rangle$ follow from crossing symmetry and reflection positivity. We will parametrize $\<TTT\>$ in general dimensions by $C_{T}$, $t_{2}$, and $t_{4}$, where $C_{T}$ is the central charge which appears in the 2-point function of the stress tensor. The relation between $t_{2}$, $t_{4}$ and the basis used in \cite{Osborn:1993cr} is given in appendix \ref{sec:TensorBasis}.

\subsection{Crossing Symmetry}
\label{subsec:CrossingTTff}
The 4-point function can be expanded in three different OPE channels. The s- and t- channel are: 
\begin{align}
\text{s-channel: \qquad} &G_T^{\mu\nu\rho\sigma}(z,\bar{z})=(z\bar{z})^{-\frac{1}{2}(\Delta_{\f}-\Delta_{T})}\sum_{\mathcal{O}}\lambda_{T\f\mathcal{O}}\lambda_{\f T\mathcal{O}}g^{\Delta_{T\f},\Delta_{\f T},\mu\nu\rho\sigma}_{\mathcal{O}}(z,\bar{z}), \label{eq:TTOOsch} \\
&\hspace{2cm}= G^{\mu\nu\rho\sigma}_{T,STT}(z,\bar{z})+ G^{\mu\nu\rho\sigma}_{T,A}(z,\bar{z})+ G^{\mu\nu\rho\sigma}_{T,B}(z,\bar{z}),
\\
\text{t-channel: \qquad} &G_T^{\mu\nu\rho\sigma}(z,\bar{z})=[(1-z)(1-\bar{z})]^{-\Delta_{\f}}\sum_{\mathcal{O},b}\lambda_{TT\mathcal{O}}^{b}\lambda_{\f \f \mathcal{O}}g^{0,0,\mu\nu\rho\sigma}_{\mathcal{O},b}(1-z,1-\bar{z}),\label{eq:TTOOtch}
\end{align}
where $\Delta_{T}=d$. We've absorbed the tensor structures into the conformal blocks, since in our configuration they become functions of $z$ and $\bar{z}$. 
The u-channel is similar to the s-channel but we will not need it explicitly.
In general $d\ge4$, there are three types of operators in the $T\times\phi$ OPE, each with a unique 3-point structure: symmetric traceless tensors $STT = \{\mathcal{O}_{[\ell]},\ell\ge0\}$, tensors $A = \{\mathcal{O}_{[\ell,1]}, \ell\ge1\}$ with a pair of indices antisymmetrized and the other $(\ell - 1)$ indices symmetrized, and tensors $B = \{\mathcal{O}_{[\ell,2]}, \ell\ge2\}$ two pairs of indices antisymmetrized and the other $(\ell - 2)$ indices symmetrized.

When $d=3$ we once again assume the theory preserves parity and use the label $STT$ for operators of even parity and the label $A$ for operators of odd parity. Finally, we use the index $b$ in (\ref{eq:TTOOtch}) to take into account that operators with spin in the $T\times T$ OPE can in general have three different parity-preserving 3-point structures in $d\geq4$ and two in $d=3$. Scalar operators appear with a unique tensor structure in the $T\times T$ OPE.

Similar to what happened for $\<J\f\f J\>$, we show in appendix \ref{sec:SpinningBlocks} that at the leading order in the lightcone limit, the s-channel expansion has the following structure:
\begin{align}\label{eq:TOOTTriangularStruct}
&G_T^{++++} = G_{T,STT}^{++++},\\
&G_T^{+3+3} = G_{T,STT}^{+3+3} + G_{T,A}^{+3+3},\\
&G_T^{34} = G_{T,STT}^{34} + G_{T,A}^{34}+ G_{T,B}^{34},
\end{align}
where in the last line we consider the function $G^{34}_{T}\equiv\frac{1}{2}\<(T^{33}-T^{44})\f\f(T^{33}-T^{44})\>$ so that we can ignore all trace terms in the 4-point function tensor structures, simplifying the analysis.

As demonstrated in section \ref{sec:JJT}, the strongest bounds follow from the positivity of individual classes of s-channel operators. In particular, we will find that analyticity and reflection positivity of $G_{T,STT}^{++++}$, $G_{T,A}^{+3+3}$, and $G_{T,B}^{34}$ on the second sheet imply the 3 conformal collider bounds on the 3 coefficients of $\<TTT\>$. 

\subsection{$\langle T^{++} \phi\phi T^{++} \rangle$}
\label{subsec:T++ffT++}
The t-channel conformal block of the stress energy tensor can be used to compute the following normalized correlation function on the second sheet: 
\begin{align}\label{eq:G++++STT}
(\eta\s^{2})^{\Delta_{\f}}&G_{T}^{++++}((1+\s)e^{-2\pi i},1+\eta\s) \nonumber\\
&= (\eta\s^{2})^{\Delta_{\f}}G_{T,STT}^{++++}((1+\s)e^{-2\pi i},1+\eta\s) \nonumber\\
&=4C_{T} \ + i \lambda_{\f\f T}\frac{ \sqrt{C_{T}} 2^d \pi ^{\frac{1}{2}-\frac{d}{2}} (d-2) (d+4) \Gamma \left(\frac{d+3}{2}\right) \Gamma (d+3) }{\left(d^2-1\right)^3  \Gamma \left(\frac{d}{2}+1\right)^2}  \\
&\qquad\qquad\qquad \times \left[(d+1) ((d-3) t_{2}+d-1)+((d-1) d-4) t_{4}\right] \frac{\eta ^{\frac{d}{2}-1}}{\sigma}
+\ldots.\nonumber 
\end{align}
Applying a contour integral as in (\ref{eq:ContourScalar}), reflection positivity and analyticity of $G_{T}^{++++}$ implies the bound
\begin{align}
\label{eq:TTTBound1}
\bigg(1-\frac{1}{d-1}t_{2}-\frac{2}{d^{2}-1}t_{4}\bigg)+\frac{d-2}{d-1}(t_{2}+t_{4})\geq0 .
\end{align}

\subsection{$\langle T^{+t}  \phi\phi T^{+t} \rangle$}
\label{subsec:T+tT+tff}
To isolate the $G_{T,A}^{+3+3}$ contribution to this correlator, we need to subtract the contribution from the $STT$ operators. This is straightforwardly done by relating $G_{T,STT}^{+3+3}$ to $G_{T,STT}^{++++}$ by conformal symmetry (see table \ref{tab:ratios}). Using the t-channel conformal block to compute $G_{T}^{+3+3}$ and subtracting the $STT$ contribution using (\ref{eq:G++++STT}), we obtain: 
\begin{align}
&(\eta\s^{2})^{\Delta_{\f}}G_{T,A}^{+3,+3}((1+\s)e^{-2\pi i},1+\eta\s)=\frac{1-d}{1+d}C_{T}\nonumber \\ & -i \lambda_{\f\f T}\frac{\sqrt{C_{T}} 2^{d-3} \pi ^{\frac{1}{2}-\frac{d}{2}}  \Gamma \left(\frac{d-1}{2}\right) \Gamma (d+3) \left[(d+1) (d (t_{2}+2)-3 t_{2}-2)-4 t_{4}\right]}{(d+1)^2  \Gamma \left(\frac{d}{2}+1\right)^2} \frac{\eta ^{\frac{d}{2}-1}}{\sigma} .\label{eq:G+t+tA}
\end{align}
Applying a contour integral as in (\ref{eq:ContourScalar}), reflection positivity and analyticity of $G_{T,A}^{+3+3}$ implies the bound\footnote{Formally, we have also removed the contribution from $B$ operators. They can only show up at subleading orders in the lightcone limit so they do not change (\ref{eq:G+t+tA}).}
\begin{align}
\label{eq:TTTBound2}
\bigg(1-\frac{1}{d-1}t_{2}-\frac{2}{d^{2}-1}t_{4}\bigg)+\frac{1}{2}t_{2}\geq0 .
\end{align}

\subsection{$\langle T^{tt}  \phi\phi T^{tt} \rangle$}
\begin{figure}
\centering
\includegraphics[width=100mm]{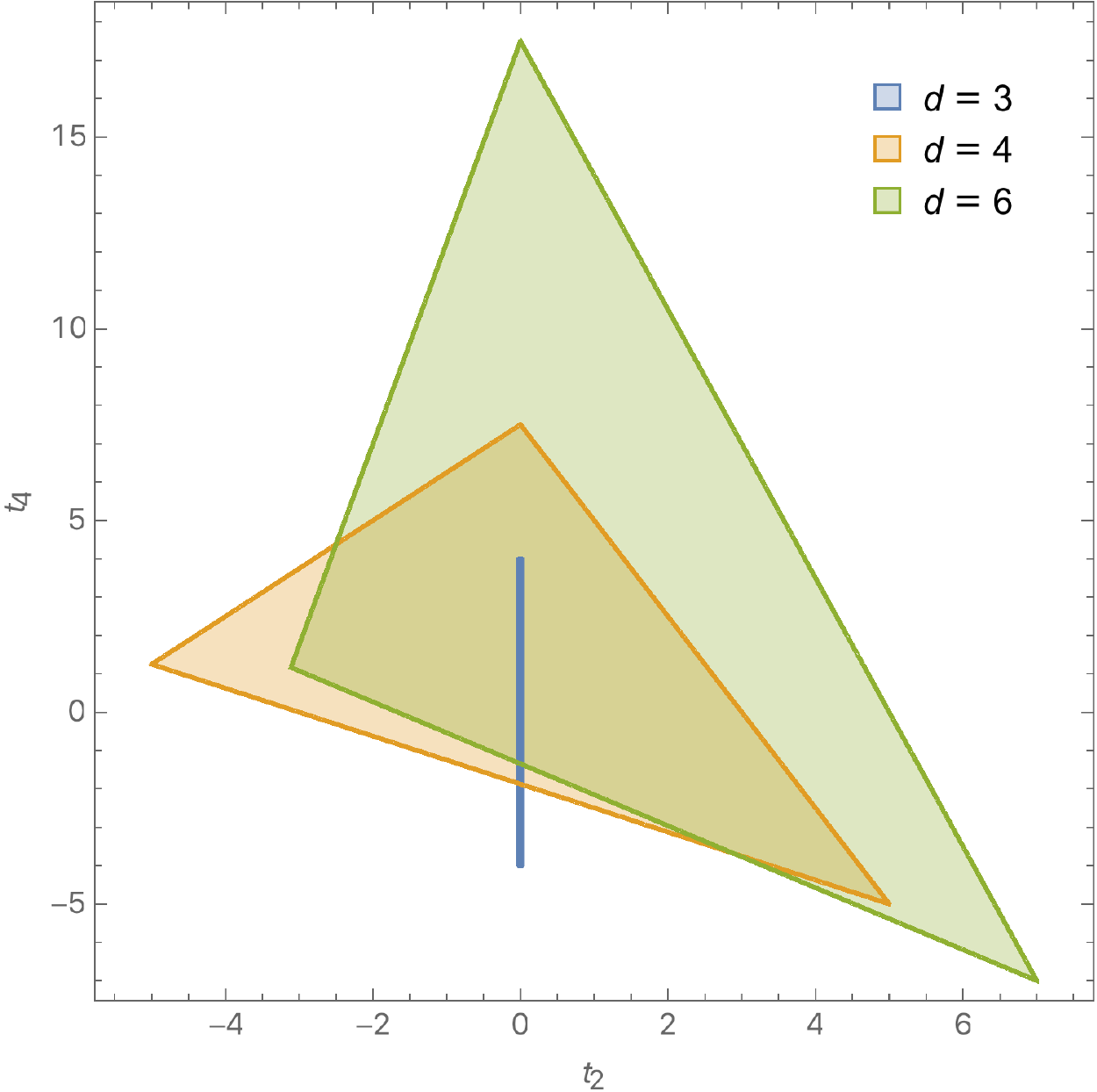}
\caption{\small Conformal collider bounds in several dimensions $d$. Filled regions correspond to the allowed values of the parameters  $t_{2}$, $t_{4}$. \label{fig:boundsplot}}
\end{figure}

 To get the exclusive contribution of the $[\ell,2]$ operators to this correlator, we need to substract the contributions from $STT$ and $A$. This is done by relating $G_{T,STT}^{34}$ to $G_{T,STT}^{++++}$ and $G_{T,A}^{34}$ to $G_{T,A}^{+3+3}$ by conformal symmetry (see table \ref{tab:ratios}). Using the t-channel conformal block of the stress tensor, (\ref{eq:G++++STT}), and (\ref{eq:G+t+tA}), we obtain: 
\label{subsec:TttTttff}
\begin{align}
(\eta\s^{2})^{\Delta_{\f}}G_{T,B}^{34}(&(1+\s)e^{-2\pi i},1+\eta\s)=\frac{(d-1)^2}{d (d+1)}C_{T} \nonumber \\& +i\lambda_{\f\f T}\frac{\sqrt{C_{T}} 2^{2 d+1} \pi ^{-\frac{d}{2}} \Gamma \left(\frac{d+1}{2}\right)^2 \left[(d+1) (d-t_{2}-1)-2 t_{4}\right]}{(d+1)  \Gamma \left(\frac{d}{2}+1\right)} \frac{\eta ^{\frac{d}{2}-1}}{\sigma}.
\end{align}
Through a contour integral similar to (\ref{eq:ContourScalar}), reflection positivity and analyticity of $G_{T,B}^{34}$ implies the bound
\begin{align}
\label{eq:TTTBound3}
\left(1-\frac{1}{d-1}t_{2}-\frac{2}{d^{2}-1}t_{4}\right)\geq0 .
\end{align}
Each of the bounds~(\ref{eq:TTTBound1}), (\ref{eq:TTTBound2}), and (\ref{eq:TTTBound3}) corresponds to a conformal collider bound in general dimensions as can be seen by comparing with \cite{Camanho:2009vw,Buchel:2009sk}. Furthermore, in $d=3$, due to the extra degeneracy in tensor structures $t_{2}=0$, and the third bound becomes equivalent to the second. Alternatively one can work in $d=3$ directly and use the first two crossing equation to rederive the $d=3$ collider bounds. In figure \ref{fig:boundsplot} we illustrate the bounds in $t_{2}$, $t_{4}$ in a few different dimensions.

These bounds are also strengthened in the presence of SUSY. In a 4d $\mathcal{N}=1$ SCFT we have the relations
\bea
t_{2}=6\left(1-\frac{a}{c}\right), \ \ t_{4}=0 \quad \Rightarrow \quad \frac{3}{2}\geq\frac{a}{c}\geq\frac{1}{2},
\eea
which yields a stronger lower bound than what we obtained from applying the $\<JJT\>$ bound to $U(1)_R$ currents, as pointed out earlier. 

In a 4d $\mathcal{N}=2$ SCFTs the lower bound is identical to the one in (\ref{eq:SUSY2ac}) but the upper bound is weaker \cite{Hofman:2009ug}, so there are no new constraints from $\<TTT\>$. These bounds are also trivially satisfied in a 4d $\mathcal{N}=4$ SCFT where $a=c$.

\subsection{Anomalous Dimensions}
\label{subsec:JfAnomalousDimT}

In the lightcone limit, the s-channel sum is dominated by the large spin double-twist operators. The large spin $STT$ operators have the following schematic form: $[T\f]^{[\ell]}_{n,\ell}=T^{\mu_1\mu_2} \partial^{(\mu_3}\dots\partial^{\mu_\ell)}\partial^{2n}\phi$. The large spin A operators $[T\f]^{[\ell,1]}_{n,\ell}$ has one index on $T$ antisymmetrized with a $\partial^\mu$, while the large spin B operators $[T\f]^{[\ell,2]}_{n,\ell}$ have both indices on $T$ antisymmetrized with $\partial^\mu$'s. We will now show that the large $\ell$ anomalous dimension asymptotics for $[T\f]^{[\ell]}_{0,\ell}$, $[T\f]^{[\ell,1]}_{0,\ell}$, and $[T\f]^{[\ell,2]}_{0,\ell}$ are negative. 

Matching the t-channel identity contribution in the s-channel yields the large $\ell$ asymptotics of the double twist OPE coefficients:
\begin{align}
(\lambda_{T\f[T\f]^{[\ell]}_{0,\ell}})^{2}&=C_{T}\frac{\sqrt{\pi } 2^{-\Delta_{\f}-d+6}}{\Gamma (\Delta_{\f}) \Gamma (d+2)}2^{-\ell}\ell^{\frac{1}{2}(-7 + 2 \Delta_{\f} + 2d)},
\\
(\lambda_{T\f[T\f]^{[\ell,1]}_{0,\ell}})^{2}&=C_{T}\frac{\sqrt{\pi } (d-1) 2^{-\Delta_{\f}-d+5}}{\Gamma (\Delta_{\f}) \Gamma (d+2)}2^{-\ell}\ell^{\frac{1}{2} (2 \Delta_{\f}+2d-5)},
\\
(\lambda_{T\f[T\f]^{[\ell,2]}_{0,\ell}})^{2}&=C_{T}\frac{\sqrt{\pi } d (d-1) 2^{-\Delta_{\f}-d+2}}{\Gamma (\Delta_{\f}) \Gamma (d+2)}2^{-\ell}\ell^{\frac{1}{2} (2 \Delta_{\f}+2d-3)}.
\end{align}

Matching the leading $\log(z)$ terms, we obtain the large spin anomalous dimensions due to the t-channel exchange of the stress energy tensor:
\begin{align}
&\gamma_{[T\f]^{[\ell]}_{0,\ell}}=\lambda_{\f\f T}\frac{2^{2 d-5} (d-2) \pi ^{-\frac{d}{2}-1} \Gamma \left(\frac{d-1}{2}\right)^2 \Gamma (\Delta_{\f}) }{\sqrt{C_{T}} (d-1) \Gamma \left(-\frac{d}{2}+\Delta_{\f}+1\right)}\notag\\
&\hphantom{\gamma_{T\f[T\f]^{[\ell]}_{0,\ell}}=\lambda_{\f\f T}2^{2 d-5} (d-2)}
\times \left[(d+1) ((d-3) t_{2}+d-1)+((d-1) d-4) t_{4}\right]\frac{1}{\ell^{d-2}},
\\
&\gamma_{[T\f]^{[\ell,1]}_{0,\ell}}=\lambda_{\f\f T}\frac{2^{2 d-5} \pi ^{-\frac{d}{2}-1} \Gamma \left(\frac{d-1}{2}\right)^2 \Gamma (\Delta_{\f}) \left[(d+1) (d (t_{2}+2)-3 t_{2}-2)-4 t_{4}\right]}{\sqrt{C_{T}} \Gamma \left(-\frac{d}{2}+\Delta_{\f}+1\right)}\frac{1}{\ell^{d-2}},
\\
&\gamma_{[T\f]^{[\ell,2]}_{0,\ell}}=\lambda_{\f\f T}\frac{2^{2 d-3} \pi ^{-\frac{d}{2}-1} \Gamma \left(\frac{d-1}{2}\right)^2 \Gamma (\Delta_{\f}) \left[(d+1) (d-t_{2}-1)-2 t_{4}\right]}{\sqrt{C_{T}} \Gamma \left(-\frac{d}{2}+\Delta_{\f}+1\right)}\frac{1}{\ell^{d-2}}.
\end{align}
Because of (\ref{eq:TTTBound1}), (\ref{eq:TTTBound2}), (\ref{eq:TTTBound3}), and $\lambda_{\phi\phi T}<0$ due to the Ward identity, these anomalous dimensions are negative. We find that this behavior also extends to $n>0$ in the $\ell\gg n$ limit (see appendix \ref{sec:NonzeroN}). For a quantum gravity theory in AdS dual to a unitary CFT, we have proven that gravity must be attractive in AdS between a scalar particle and a graviton separated by super-horizon distances. Once again, this result does not make any assumption about being in a large $N$ limit. 

\section{Non-Conserved Currents}
\label{sec:NonConserved}
We will now generalize the above discussion to external non-conserved operators. In particular we will be interested in the correlation function $G^{\mu\nu}_{V}(z,\bar{z})=\<V^{\mu}(0)\f(z,\bar{z})\f(1) V^{\nu}(\infty)\>$ where $V^{\mu}$ is a generic vector with dimension $\Delta_{V} > d-1$ and $\f$ is a scalar of arbitrary dimension. In \cite{Komargodski:2016gci} it was observed that the bounds derived via deep inelastic scattering (DIS) are weaker than the bounds derived from positivity of the energy one point function $\<\mathcal{E}(\vec{n})\>$. Here we will find that the bounds we obtain from reflection positivity and crossing symmetry for this correlation function coincide with the results of the DIS argument.

The general 3-point function $\<V\f\mathcal{O}\>$, where $\mathcal{O}$ is a symmetric traceless tensor primary, using the formalism of \cite{Costa:2011mg,Costa:2011dw}, takes the form
\bea
(b_{1}D_{11}\Sigma^{1,0}+b_{2}D_{12}\Sigma^{0,1})\frac{V_{3}^{\ell}}{P_{12}^{\frac{1}{2}(\Delta_{V}+\Delta_{\f}-\Delta_{\mathcal{O}}-\ell)}P_{13}^{\frac{1}{2}(\Delta_{V}+\Delta_{\mathcal{O}}+\ell-\Delta_{\f})}P_{23}^{\frac{1}{2}(\Delta_{\f}+\Delta_{\mathcal{O}}+\ell-\Delta_{V})}}.
\eea

When $\mathcal{O}$ is in the $[\ell,1]$ representation of $\SO(d)$ for $d\geq4$, or a parity odd operator in $d=3$, the 3-point function $\<V\f\mathcal{O}\>$ is uniquely determined up to a single overall OPE coefficient. The corresponding conformal block is the same as when $V$ is a conserved current and is written in appendix \ref{ap:SeedBlockA}

Matching the identity block in the t-channel we find that the double-twist operators $[V\f]^{[\ell]}_{n,\ell}$ and $[V\f]^{[\ell,1]}_{n,\ell}$ with twist $\tau_{V}+\Delta_{\f}+2n$ and $\tau_{V}+1+\Delta_{\f}+2n$, respectively, must be present in the regime $\ell\gg n$. Moreover the effect of the $b_{1}$ term is necessarily subleading at small $(1-\bar{z})$. We then define the OPE coefficient $\lambda_{V\f[V\f]^{[\ell]}_{n,\ell}}$ to be equal to $b_{2}$. We will also canonically normalize the 2-point function as 
\bea
\<V^{\mu}(x)V^{\nu}(0)\>=\frac{I^{\mu\nu}(x)}{x_{12}^{2\Delta_{V}}}.
\eea

Once again, we find that in reproducing the divergences in the t-channel, the $STT$ operators will contribute to both the $++$ crossing equation and the $tt$ crossing equation, while the $[\ell,1]$ operators will only contribute to the $tt$ equation. The ratio between the $STT$ contribution in the $tt$ polarization and the $++$ polarization is $\frac{-1}{2\Delta_{V}}$ for the identity matching and $-\frac{1}{2}\frac{1}{1+\Delta_{V}-d/2}$ for the stress tensor matching.

Next, we need to find the general form of the 3-point function $\<VVT\>$. Working in the differential basis and imposing conservation we obtain: 
\bea
\<VVT\>= \bigg(e_{1}D_{11}D_{22}+e_{2}D_{12}D_{21}+e_{3}H_{12}\bigg)\Sigma^{1,1}\frac{V_{3}^{2}}{P_{12}^{\Delta_{V}-d/2-1}P_{13}^{d/2+1}P_{23}^{d/2+1}}.
\eea
The relation between this basis and the basis used in \cite{Komargodski:2016gci} is
\begin{align}
e_{1}=a_{3}, \qquad e_{2}=-2a_{2}-a_{3}, \qquad e_{3}=a_{1}-2(d/2-1)a_{2}+a_{3}(1-d),
\end{align}
where the Ward identity additionally imposes the condition
\begin{align}
a_{1}=-(\Delta_{V}-d+1)(a_{2}+a_{3}).
\end{align}

Similar to the case of conserved vectors, we will be interested in the following two functions:
\begin{align}
&G^{++}_{V}(z,\bar{z})\equiv \<V^{+}(0)\f(z,\bar{z})\f(1) V^{+}(\infty)\>,
\\
&G^{tt}_{V,A}(z,\bar{z})\equiv G^{tt}_{V}(z,\bar{z})-G^{tt}_{V,STT}(z,\bar{z}).\end{align}

Going to the second sheet $z\rightarrow ze^{-2\pi i}$, writing $z=1+\s$ and $\bar{z}=1+\eta\s$, and going to the limit $\eta\ll |\s| \ll1$, we find
\begin{align}
&(\eta\s^{2})^{\Delta_{\f}}G^{++}_{V}((1+\s)e^{-2\pi i},1+\eta\s) = -2\nonumber\\&- i\lambda_{\f\f T}\frac{2^{2 d+1}  \Gamma \left(\frac{d+1}{2}\right) \Gamma \left(\frac{d+3}{2}\right) \left[a_{2} \left(d^2-6 d+4 \Delta_{V}+4\right)+4 a_{3} (-d+\Delta_{V}+1)\right]}{ \sqrt{C_{T}}  \Gamma \left(\frac{d}{2}+1\right)^2} \frac{\eta ^{\frac{d}{2}-1}}{\sigma},\\
&(\eta\s^{2})^{\Delta_{\f}}G^{tt}_{V,A}((1+\s)e^{-2\pi i},1+\eta\s)=\frac{\Delta_{V}-1}{\Delta_{V}}\notag\\
&+i \lambda_{\f\f T}\frac{4^{d+1}  \Gamma \left(\frac{d+1}{2}\right) \Gamma \left(\frac{d+3}{2}\right) \left[a_{3} \left(d (\Delta_{V}-2)-2 \Delta_{V}^2+2\right)+2 (\Delta_{V}-1) a_{2} (d-\Delta_{V}-1)\right]}{ \sqrt{C_{T}}  (d-2 (\Delta_{V}+1)) \Gamma \left(\frac{d}{2}+1\right)^2} \frac{\eta ^{\frac{d}{2}-1}}{\sigma}.
\end{align}

Using same contour integral argument as in the previous sections we find the bounds
\begin{align}
a_{2}\leq 0 \ \ \& \ \ a_{3}\geq&\frac{a_{2}\left(d^2-6 d+4 \Delta_{V}+4\right)}{4 (d-\Delta_{V}-1)},
\\
&\text{or} \nonumber
\\
a_{2}>0 \ \ \& \ \ a_{3}\geq&\frac{2a_{2} \left(-d \Delta_{V}+d+\Delta_{V}^2-1\right)}{d (\Delta_{V}-2)-2 \Delta_{V}^2+2}.
\end{align}

These constraints match exactly on to the first line of (4.28) and (4.29) of \cite{Komargodski:2016gci}, while positivity of the one point energy function yields a stronger constraint when $a_{2}$ is less than zero. 

Finally, we will repeat the solving of the lightcone bootstrap equation on the first sheet. Matching the identity contribution, we find the $n=0$ asymptotic behavior for the OPE coefficients
\bea
(\lambda_{V\f[V\f]^{[\ell]}_{0,\ell}})^{2}=\frac{\sqrt{\pi } 2^{-\Delta_{V}-\Delta_{\f}+4}}{\Gamma (\Delta_{V}+1) \Gamma (\Delta_{\f})}\ell^{\frac12 (2 \Delta_{V} + 2\Delta_{\f}-5)},
\\
(\lambda_{V\f[V\f]^{[\ell,1]}_{0,\ell}})^{2}=\frac{\sqrt{\pi } (\Delta_{V}-1) 2^{-\Delta_{V}-\Delta_{\f}+2}}{\Gamma (\Delta_{V}+1) \Gamma (\Delta_{\f})}\ell^{\frac{1}{2} (2 \Delta_{V}+2 \Delta_{\f}-3)}.
\eea 

The large spin anomalous dimension asymptotics due to the exchange of the stress tensor for the $n=0$ double twist states are given by:
\begin{align}
&\gamma^{[\ell]}_{0,\ell}=\lambda_{\f\f T}\frac{\Gamma (d+2) \Gamma (\Delta_{\f}) \Gamma (\Delta_{V}+1) \left[a_{2} \left(d^2-6 d+4 \Delta_{V}+4\right)+4 a_{3} (-d+\Delta_{V}+1)\right]}{4 \sqrt{C_{T}} \Gamma \left(\frac{d}{2}+1\right)^2 \Gamma \left(-\frac{d}{2}+\Delta_{\f}+1\right) \Gamma \left(-\frac{d}{2}+\Delta_{V}+2\right)}\frac{1}{\ell^{d-2}},
\\
&\gamma^{[\ell,1]}_{0,\ell}=\lambda_{\f\f T}\frac{\Gamma (d+2) \Gamma (\Delta_{\f}) \Gamma (\Delta_{V}+1) }{2 \sqrt{C_{T}} (\Delta_{V}-1) \Gamma \left(\frac{d}{2}+1\right)^2 \Gamma \left(-\frac{d}{2}+\Delta_{\f}+1\right) \Gamma \left(-\frac{d}{2}+\Delta_{V}+2\right)}\notag\\
&\hphantom{\gamma^{[\ell,1]}_{0,\ell}=\lambda_{\f\f T}}
\times\left[2 a_{2} \left(-d \Delta_{V}+d+\Delta_{V}^2-1\right)+a_{3} \left(-d \Delta_{V}+2 d+2 \Delta_{V}^2-2\right)\right]\frac{1}{\ell^{d-2}}.
\end{align}
These are negative if and only if the the constraints derived above hold.

\section{Discussion}
\label{sec:discussion}

In this work, we have proven that the ``conformal collider bounds" originally proposed in~\cite{Hofman:2008ar} hold for any unitary parity-preserving conformal field theory (CFT) with a unique stress tensor in dimensions $d\ge 3$. This presents the first complete field theory proof of the ``conformal collider bounds" conjectured in \cite{Hofman:2008ar}. While there was a large amount of evidence suggesting the result was indeed correct, as reviewed in the introduction, until now it had remained remarkably difficult to obtain a full proof purely based on unitarity, conformal symmetry, and quantum field theory axioms. Due to the ubiquitous relevance of the energy momentum and current 3-point functions and their relation to anomalies, it is reassuring to find that the basic principles of conformal field theory imply these constraints.

Our result also shows the power of the bootstrap methods, first championed by \cite{Ferrara:1973yt,Polyakov:1974gs,Belavin:1984vu} and revived more recently in \cite{Rattazzi:2008pe}. In its strongest form, unitarity and crossing symmetry may contain all necessary information to classify the whole landscape of conformal field theories without any extra input. Numerical results in recent years certainly point in this direction (e.g. \cite{ElShowk:2012ht,El-Showk:2014dwa,Kos:2014bka,Simmons-Duffin:2015qma,Poland:2011ey,Beem:2013qxa,Nakayama:2014lva,Chester:2014fya,Beem:2014zpa,Bobev:2015vsa,Bobev:2015jxa,Chester:2015qca,Iliesiu:2015qra,Kos:2015mba,Beem:2015aoa,Poland:2015mta}). 
If it is possible to define all CFTs through the bootstrap method, it would be a significant step forward in the understanding of strongly coupled quantum field theory as they are connected to CFTs by renormalization group flows.
This work adds to the list of positive results by tackling a problem that had resisted attacks from other methods.

In parallel to the numerical approach, there is an important line of research that uses analytic methods to explore the consequences of the crossing symmetry~\cite{Fitzpatrick:2012yx,Komargodski:2012ek,Fitzpatrick:2014vua,Kaviraj:2015cxa,Alday:2015ota,Vos:2014pqa,Alday:2015eya,Kaviraj:2015xsa,Fitzpatrick:2015qma,Li:2015rfa,Alday:2015ewa,Li:2015itl,Dey:2016zbg}. This approach has opened up regimes of the CFT data that are difficult to probe numerically, but which have clear and crucial relevance to quantum gravitational theories in AdS. For example, AdS observables such as binding energies~\cite{Fitzpatrick:2012yx,Komargodski:2012ek,Li:2015rfa,Li:2015itl} and Eikonal phases~\cite{Cornalba:2006xm,Cornalba:2006xk,Cornalba:2007zb} are directly related to the anomalous dimensions of classes of CFT operators with large dimensions and large spin, which are more accessible with the analytic bootstrap methods. 

More generally, once we understand CFTs, we can use them as starting points to answer important questions in quantum gravity. An especially exciting question is the quantum origin of universal features of gravitational interactions, such as causality, (non-)locality and attractiveness~\cite{Heemskerk:2009pn,Fitzpatrick:2012cg,Fitzpatrick:2012yx,Komargodski:2012ek,Fitzpatrick:2014vua,Fitzpatrick:2015zha,Maldacena:2015iua}. The bounds proved in this paper are directly related to properties of 3-particle vertices in the bulk, including at least one graviton \cite{Hofman:2008ar}. These bounds also imply that the gravitational interaction is attractive between two particles separated by super-horizon distances in AdS, as a direct consequence of the unitarity of the underlying quantum theory. It would be interesting to apply the bootstrap method to other questions in quantum gravity, for instance: finding all CFTs dual to approximate Einstein gravity in AdS, where $a=c$. It was shown in \cite{Camanho:2014apa} using bulk causality that $|\frac{a-c}{c}|$ satisfies a bound directly related to the dimension of the lightest higher spin single particle operator. Establishing this result within the CFT would be a solid step towards understanding the most general UV completions of Einstein gravity in AdS.

Another important direction is to prove a much stronger positive energy condition from CFT principles. It is possible that unitarity and consistency require the energy flux operators constructed in \cite{Hofman:2008ar} to have only positive semi-definite eigenvalues, so they acquire positive expectation values in any state, not just the ones created by conserved local operators. This result is not only important theoretically, but is also relevant in experimental setups where energy correlation functions are used because of their IR finiteness properties. The obvious first step is to understand the conformal collider bounds involving non-conserved currents. The results in section \ref{sec:NonConserved}, being the same as the corresponding bounds obtained from the Deep Inelastic Scattering approach in \cite{Komargodski:2016gci}, are weaker than the conformal collider bounds and still allow for negative energy flux. It would be interesting to see whether the full energy flux positivity conditions can be derived by considering more complicated 4-point functions.

In this work we have assumed that the stress tensor is the unique conserved, spin-2 operator.
One may ask what happens if there are many such operators in the spectrum. If the theory is a product of CFTs, each with a unique stress tensor, then our arguments will go through for each sector of the theory.\footnote{The result of \cite{Alba:2015upa,Maldacena:2011jn} also holds for each sector and the conformal collider bounds are satisfied where there is higher spin symmetry in a subset of the decoupled sectors.} In general, the number of conserved, spin-2 operators does not necessarily correspond to the number of mutually decoupled sectors in the theory \cite{Maldacena:2011jn}.
Another simple example is a CFT with non-Abelian global symmetries. The current OPE $J^a \times J^b$ may contain different conserved spin-2 currents in different representations. However, as long as the stress tensor is the unique singlet spin-2 conserved current, our argument applies after projecting this OPE to the singlet sector. 
In more general cases, direct application of our method leads to bounds on certain linear combinations of OPE coefficients involving multiple spin-2 conserved currents. It is interesting to see whether there is a method that selects the stress tensor from this group, leading to the conformal collider bounds for generic CFTs with multiple spin-2 conserved currents. 

Last but not least, efforts in the modern conformal bootstrap program have largely focused on the 4-point function of scalar operators. This is primarily due to limited knowledge of spinning conformal blocks. However, it was already demonstrated in \cite{Iliesiu:2015qra, Li:2015itl, Hartman:2016dxc} and this work that the crossing symmetry of 4-point functions involving spinning operators harbors rich information inaccessible from scalar 4-point functions. Fortunately, much progress has taken place in the last year on spinning conformal blocks. For example, the work in \cite{Costa:2011mg, Costa:2011dw, SimmonsDuffin:2012uy, Rejon-Barrera:2015bpa, Li:2015itl, Costa:2014rya} was of crucial importance to develop the results in this paper. Other important progress on spinning blocks includes~\cite{Iliesiu:2015akf,Echeverri:2015rwa,Echeverri:2016dun}. These results are extremely valuable and could lead to new breakthroughs for the bootstrap, both on the numerical and analytical front. Extending these ideas to include the energy-momentum tensor would also open up new opportunities to understand universal physics in CFTs and the bulk properties of their holographic duals.

\section*{Acknowledgements}
We thank Tom Hartman, Jared Kaplan, Daniel Robbins, Slava Rychkov, David Simmons-Duffin, and Junpu Wang for discussions. The work of DP and DM is supported by NSF grant PHY-1350180. DP is additionally supported as a Martin A. and Helen Chooljian Founders' Circle Member at IAS. The work of DL is supported in part by NSF grants PHY-1316665 and PHY-1454083. FR is supported by the Mexican Consejo Nacional de Ciencia y Tecnolog\'{i}a (CONACyT scholarship 382219).

\appendix
\section{Spinning Conformal Blocks at Large Spin}
\label{sec:SpinningBlocks}
In this appendix we derive the relevant s-channel spinning conformal blocks in the lightcone limit. We find that they can be written as (derivatives of) a single scalar conformal block.  This simplification occurs because the sum over spins in the s-channel is performed via 
\begin{align}\label{eq:sumoverlint}
 \int_{0}^{\infty}d\ell~\ell^{\alpha}K_{\nu}(2\ell \sqrt{1-\bar{z}})=\frac{(1-\bar{z})^{-(\alpha+1)/2}}{4}\Gamma \left( \frac{1+\alpha- \nu}{2} \right) 
 \Gamma \left( \frac{1+\alpha+\nu}{2} \right),
 \end{align} 
 so that terms with higher powers in $1/\ell$ result in subleading terms in $(1-\bar{z})$, for $(1-\bar{z})\ll 1$ and $\ell\gg 1$ with $(1-\bar{z})\ell^{2}\lesssim 1$. Therefore the idea is to count the relative powers of $1/\ell$, where $(1-\bar{z})$ has a weight of $O(1/\ell^{2})$. 

Our strategy will be to write the conformal blocks as differential operators acting on a basic set of `seed' blocks, following the general approach developed in~\cite{Costa:2011dw,Costa:2014rya,Iliesiu:2015qra,Rejon-Barrera:2015bpa,Echeverri:2015rwa,Iliesiu:2015akf,Echeverri:2016dun}. For the seed blocks we expect, motivated by the results in \cite{Rejon-Barrera:2015bpa}, that they can be written as $g_{\mathrm{seed}}\sim g_{\mathrm{scalar}}+(\dots)$, where the term $(\dots)$ includes $STT$ conformal blocks of higher-spin correlators (see for example Eq. (4.87) in that paper). Moreover one can check, using the results of \cite{Costa:2011dw}, that the spinning blocks in $(\dots)$ are sub-leading in $1/\ell$ with respect to $g_{\mathrm{scalar}}$.\footnote{The differential operators are constructed in such a way that the spin is increased while maintaining the dimensions of the original three-point function. Thus there are no relative powers of $(1-\bar{z})$ coming from the difference in external dimensions (see (\ref{eq:f1def})). The sub-leading powers of $\ell$ then come from the matrix transforming the differential basis to the standard one.} For blocks that can be derived from seed blocks, the simplifications can be inferred by looking at the differential operators of \cite{Costa:2011dw} when acting on seeds.

The projection of these results into different polarizations gives an explicit check of the the triangular structure (\ref{eq:TOOTTriangularStruct}) for finite $z$. Furthermore, contributions from each irreducible representation are related by a $z$-independent factor, at each order in $(1-\bar{z})$.

\subsection{Seed Blocks}
First we look at the seed conformal blocks for the $[\ell,1]$ and $[\ell,2]$ representations that appear in $\langle J \phi \phi J \rangle$ and $\langle T \phi \phi T \rangle$ respectively. 

 Notice that the simplification for $[\ell,1]$ can be easily obtained by taking the $\bar{z}\to 1$, $(1-\bar{z})\ell^{2}\lesssim 1$ limit in the expressions for $g_{A}$ given in (4.87)--(4.91) of \cite{Rejon-Barrera:2015bpa}. Nonetheless we include this calculation given that the logic is the same as for the $[\ell,2]$ blocks, where the explicit expressions are not known yet.
 
 \subsubsection{$[\ell,1]$ Seed} 
 \label{ap:SeedBlockA}
 The integral representation of the $[\ell,1]$ conformal block in $\langle J \phi \phi J \rangle$ is given by 
 \begin{multline}
 g_{A}^{\Delta_{J\phi},\Delta_{\phi J},\mu \nu}(z,\bar{z})
 \\=\frac{\mathcal{N}_{A}(\lambda_{\phi J \widetilde{A}}/\lambda_{\phi J A})}{X_{\Delta_{i}}}
 \int d^{d}x_{0}~ \langle J_{\mu}(x_{1})\phi(x_{2})A(x_{0}) \rangle
 \Pi^{[\ell,1]} 
 \langle \widetilde{A}(x_{0})\phi(x_{3}) J_{\nu}(x_{4}) \rangle,
 \end{multline}
 where the tensor contraction is $\Pi^{[\ell,1]} = m^{(10)}_{\mu \rho}\mathcal{P}^{[\ell,1]\, \rho}_{\sigma}m^{(40)\, \sigma}_{ \nu}$, with 
 \begin{align}
 \mathcal{P}^{[\ell,1]\, \rho}_{\sigma}(k^{(012)},k^{(034)})\equiv
 k^{(012)}_{\rho_{1}}\cdots k^{(012)}_{\rho_{\ell}}\Pi^{[\ell,1]\, \rho \rho_{1}\cdots \rho_{\ell}}_{\sigma \sigma_{1}\cdots \sigma_{\ell}}k^{(034)\, \sigma_{1}}\cdots k^{(034)\, \sigma_{\ell}}.
 \end{align}
 Here $k$ and $m$ are given in (A.3) and (A.4) of \cite{Rejon-Barrera:2015bpa} respectively, $\tilde{A}$ is a shadow operator, and the integral has an implicit monodromy projection (as discussed in~\cite{SimmonsDuffin:2012uy}).
 Using the results of \cite{Eilers:2006kd} we can write this tensor as
 \begin{align}\label{eq:mixedtensorcontraction}
 \mathcal{P}_{\sigma}^{[\ell,1]\, \rho}(X,Y)=\frac{1}{\ell+1}\left( \ell\delta_{\sigma}^{\rho}
 +\frac{X^{2}\partial_{\sigma}\partial^{\rho}-(\ell-1)X^{\rho}\partial_{\sigma}}{d+\ell-3}-X_{\sigma}\partial^{\rho} \right) \mathcal{P}^{[\ell]}(X,Y),
 \end{align}
 where $\mathcal{P}^{[\ell]}$ is the traceless-symmetric contraction of $\ell$ indices,
 \begin{align}
 \mathcal{P}^{[\ell]}(X,Y)=X_{a_{1}}\cdots X_{a_{\ell}}\Pi^{[\ell]\, a_{1}\cdots a_{\ell}}_{b_{1}\cdots b_{\ell}}Y^{b_{1}}\cdots Y^{b_{\ell}}.
 \end{align}
 Notice that derivatives acting on $\mathcal{P}^{[\ell]}$ are structures that appear in $STT$ spinning blocks, and thus sub-leading with respect to scalar blocks. Therefore keeping only the first term in (\ref{eq:mixedtensorcontraction}) leads to a single scalar block times a tensor structure,
 \begin{multline}\label{eq:mixedsymmapprox}
 g_{A}^{\Delta_{J\phi},\Delta_{\phi J},\mu \nu}(z,\bar{z})\\
 =\frac{\mathcal{N}_{A}(\lambda_{\phi J \widetilde{A}}/\lambda_{\phi J A})}{\mathcal{N}_{\mathcal{O}}(\lambda_{\phi J \widetilde{\mathcal{O}}}/\lambda_{\phi J \mathcal{O}})}
 g_{(\Delta_{A},\ell)}^{\Delta_{J\phi},\Delta_{\phi J}}(z,\bar{z})
 \left( m^{(14)}_{\mu \nu}+2(z\bar{z})^{-\frac12}k^{(124)}_{\mu}k^{(413)}_{\nu} \right)+O(1/\ell),
 \end{multline}
 where the prefactor is given in (G.1) of \cite{Rejon-Barrera:2015bpa}. Notice that this prefactor can always be set to one by changing the normalization of $A$.  
Evaluating the relevant polarizations leads to 
\begin{align}
g_{A}^{\Delta_{J\phi},\Delta_{\phi J},++}(z,\bar{z})&=O((1-\bar{z})^{1}),\\
\label{eq:ablockdef}
g_{A}^{\Delta_{J\phi},\Delta_{\phi J},tt}(z,\bar{z})&= g^{\Delta_{J\phi},\Delta_{\phi J}}_{\Delta_{A},\ell}(z,\bar{z})+O(1/\ell),
\end{align}
where the tensor structure of the last term is of order $O((1-\bar{z})^{0})$.
 \subsubsection{$[\ell,2]$ Seed}
For $\langle T\phi\phi T\rangle$ we have 
\begin{multline}
g_{B}^{\Delta_{T\phi},\Delta_{\phi T},\mu \nu\rho \sigma}(z,\bar{z})
\\
=\frac{\mathcal{N}_{B}(\lambda_{\phi T\widetilde{B}}/\lambda_{\phi TB})}{X_{\Delta_{i}}}\int d^{d}x_{0}~\langle T_{\mu \nu}(x_{1})\phi(x_{2})B(x_{0})\rangle \Pi^{[\ell,2]}\langle \widetilde{B}(x_{0})\phi(x_{3})T_{\rho \sigma}(x_{4})\rangle,
\end{multline}
where $B\in [\ell,2]$. Here the contraction is $\Pi^{[\ell,2]} = m^{(10)}_{\mu \alpha_{1}}m^{(10)}_{\nu \alpha_{2}}\mathcal{P}^{[\ell,2]\, \alpha_{1}\alpha_{2}}_{\beta_{1}\beta_{2}}m^{(40)\, \beta_{1}}_{\rho }m^{(40)\, \beta_{2}}_{\sigma }$, with \cite{Eilers:2006kd}
\begin{align}
\mathcal{P}^{[\ell,2]\, \alpha_{1}\alpha_{2}}_{\beta_{1}\beta_{2}}(X,Y)
=\left(\frac{\ell-1}{\ell+1}\delta^{\alpha_{1}}_{( \beta_{1}}\delta^{\alpha_{2}}_{\beta_{2})}+\mathrm{derivatives}\right)\mathcal{P}^{[\ell]}(X,Y).
\end{align}
By the same arguments as in the previous case
\begin{multline}\label{eq:bblockdef}
g_{B}^{\Delta_{T\phi},\Delta_{\phi T},\mu\nu\rho\sigma}(z,\bar{z})\\ 
=\frac{\mathcal{N}_{B}(\lambda_{\phi T\widetilde{B}}/\lambda_{\phi TB})}{\mathcal{N}_{\mathcal{O}}(\lambda_{\phi T\widetilde{\mathcal{O}}}/\lambda_{\phi T\mathcal{O}})}g^{\Delta_{T\phi},\Delta_{\phi T}}_{(\Delta_{B},\ell)}(z,\bar{z}){\Pi^{[2]}}^{\mu\nu;\alpha\beta}{\Pi^{[2]}}^{\rho\sigma;\gamma\delta}\left(
m^{(14)}_{\alpha\gamma}+2 (z\bar{z})^{-\frac12}k^{(124)}_{\alpha}k^{(413)}_{\gamma}
\right)\\\times\left(
m^{(14)}_{\beta\delta}+2 (z\bar{z})^{-\frac12}k^{(124)}_{\beta}k^{(413)}_{\delta}
\right)+O(1/\ell),
\end{multline}
Notice that we can set the prefactor to one by a suitable normalization of $B$. Evaluating the relevant polarizations at lowest order in $O(1-\bar{z})$ gives 
\begin{align}
g_{B}^{\Delta_{T\phi},\Delta_{\phi T},++++}(z,\bar{z})&=O((1-\bar{z})^{2}),\\
g_{B}^{\Delta_{T\phi},\Delta_{\phi T},+3+3}(z,\bar{z})&=O((1-\bar{z})^{1}),\\
g_{B}^{\Delta_{T\phi},\Delta_{\phi T},34}(z,\bar{z})&=  g^{\Delta_{T\phi},\Delta_{\phi T}}_{\Delta_{B},\ell}(z,\bar{z})+O(1/\ell),
\end{align}
where the last term's tensor structure is of order $O((1-\bar{z})^{0})$.

\subsection{Derived Blocks}
Now we turn to the conformal blocks that can be obtained from seeds, by acting with the differential operators $D_{ij}$ of \cite{Costa:2011dw}.\footnote{It may also be interesting to derive these blocks more directly by expressing the OPE in embedding space~\cite{Fortin:2016lmf}.} The $STT$ exchange in both $\langle J\phi\phi J \rangle$ and $\langle T\phi\phi T \rangle$ can be computed from the lightcone approximation to the scalar block \cite{Fitzpatrick:2012yx}
\begin{align}\label{eq:scalarapprox}
 g^{\Delta_{12},\Delta_{34}}_{\tau,\ell}(u,v)\overset{\substack{\ell\gg 1,~(1-\bar{z})\ell^{2}\lesssim 1\\v\ll u<1}}{=}
 f_{1}^{\Delta_{12},\Delta_{34}}(\ell,1-\bar{z})f_{2}^{\Delta_{12},\Delta_{34}}(\tau,u)(1+O(1/\sqrt{\ell},\sqrt{1-\bar{z}})),
 \end{align} 
where $v\approx (1-\bar{z})(1-u)$, $u\approx z$, and
\begin{align}\label{eq:f1def}
 f_{1}^{\Delta_{12},\Delta_{34}}(\ell,x) &=\left(-\frac12\right)^{\ell}\pi^{-\frac12}2^{2\ell}\ell^{\frac12}x^{\frac{\Delta_{12}-\Delta_{34}}{4}}K_{\frac{\Delta_{34}-\Delta_{12}}{2}}(2\ell\sqrt{x}),\\
 f_{2}^{\Delta_{12},\Delta_{34}}(\tau,u) &=\frac{2^{\tau}u^{\frac{\tau}{2}}}{(1-u)^{\frac{d}{2}-1}}{_2F_{1}}\left(\frac{\tau-d+2-\Delta_{12}}{2},\frac{\tau-d+2+\Delta_{34}}{2},\tau-d+2;u\right).
\end{align}
This limit holds for even $d\geq 2$ as long as the sum over $\ell$ only receives contributions in the region where the product $\ell^{2}(1-\bar{z})$ is kept fixed.
For the $[\ell,1]$ exchange in $\langle T\phi\phi T \rangle$ the procedure is completely analogous given that its seed is also a scalar conformal block, as shown in (\ref{eq:ablockdef}). In both cases one can analyze the differential operators and drop derivatives as well as powers of $\ell$ and $(1-\bar{z})$ that produce subleading terms. The results are summarized below.

\subsubsection{STT}
For $\langle J\phi \phi J \rangle$, the differential operator  is 
\begin{align}
\left( a_{1}^{L}D_{11}\Sigma^{1,0}_{L}+D_{12}\Sigma^{0,1}_{L} \right) 
\left( a_{1}^{R}D_{44}\Sigma^{0,1}_{R}+D_{43}\Sigma^{1,0}_{R} \right).
\end{align}
The $a_{1}^{L,R}$ terms can be found by imposing conservation, but their effect is subleading in the lightcone limit. The action of the differential operators on partial waves leads to 
\begin{align}
g^{\Delta_{J\phi},\Delta_{\phi J},++}_{\Delta_{\mathcal{O}},\ell}(u,v) &=2[v\partial_{v}-\Delta_{J\phi}]g^{\Delta_{J\phi},\Delta_{\phi J},tt}_{\Delta_{\mathcal{O}},\ell}(u,v),\\
g^{\Delta_{J\phi},\Delta_{\phi J},tt}_{\Delta_{\mathcal{O}},\ell}(u,v) &=\frac12 \sqrt{u}[\Delta_{J\phi}-1-v\partial_{v}] g^{\Delta_{J\phi}-1,\Delta_{\phi J}+1}_{\Delta_{\mathcal{O}},\ell}(u,v)(1+O(1/\ell)).
\end{align}

For $\langle T\phi\phi T\rangle$, the differential operator is 
\begin{multline}
\left( b_{1}^{L}(D_{11})^{2}\Sigma^{2,0}_{L}+b_{2}^{L}D_{12}D_{11}\Sigma^{1,1}_{L}+(D_{12})^{2}\Sigma^{0,2}_{L} \right) 
\\\times 
\left( b_{1}^{R}(D_{44})^{2}\Sigma^{0,2}_{R}+b_{2}^{R}D_{43}D_{44}\Sigma^{1,1}_{R}+(D_{43})^{2}\Sigma^{2,0}_{R} \right) .
\end{multline}
Similar to the previous case, the contribution of the $b_{1,2}^{L,R}$ terms are fixed by conservation and subleading in $1/\ell$. Counting powers in the differential operator gives
\begin{align}
g^{\Delta_{T\phi},\Delta_{\phi T},++++}_{\Delta_{\mathcal{O}},\ell}(u,v)&=2[v\partial_{v}-(\Delta_{T\phi}+1)]g^{\Delta_{T\phi},\Delta_{\phi T},+3,+3}_{\Delta_{\mathcal{O}},\ell}(u,v),\\
g^{\Delta_{T\phi},\Delta_{\phi T},+3+3}_{\Delta_{\mathcal{O}},\ell}(u,v)&=[v\partial_{v}-\Delta_{T\phi}]g^{\Delta_{T\phi},\Delta_{\phi T},34}_{\Delta_{\mathcal{O}},\ell}(u,v),\\
g^{\Delta_{T\phi},\Delta_{\phi T},34}_{\Delta_{\mathcal{O}},\ell}(u,v)&=\frac{u}{2}[(\Delta_{T\phi}-2)(\Delta_{T\phi}-1)\nonumber\\&\,\,\,\,+v(4-2\Delta_{T\phi }+v\partial_{v})\partial_{v}] g^{\Delta_{T\phi}-2,\Delta_{\phi T}+2}_{\Delta_{\mathcal{O}},\ell}(u,v)(1+O(1/\ell)).
\end{align}

\subsubsection{$[\ell,1]$}
In this case the seed 3-point function is given by (here we are using the formalism of \cite{Costa:2014rya})
\begin{align}
\langle J(P_{1};Z_{1})\phi(P_{2})A(X_{3};Z_{3},\Theta_{3})\rangle=\frac{V_{3}^{(\Theta_{3})}H_{13}^{(Z_{1},\Theta_{3})}(V_{3}^{(Z)})^{\ell-1}}{P_{12}^{\frac12(\Delta_{\phi}+\Delta_{J}-\Delta_{A}-\ell)}P_{13}^{\frac12(\Delta_{J\phi}+\Delta_{A}+\ell+2 )}P_{23}^{\frac12(\Delta_{A}-\Delta_{J\phi}+\ell)}}.
\end{align}
To construct $\langle T(P_{1};Z_{1})\phi(P_{2})A(X_{3};Z_{3},\Theta_{3}) \rangle $ we act with a linear combination of $D_{11}\Sigma^{1,0}$ and $D_{12}\Sigma^{0,1}$ and impose conservation. The spinning blocks for this exchange are then given by acting on partial waves $W_{A}$ with the differential operator
\begin{align}
\left(\lambda_{1}^{L}D_{11}\Sigma^{1,0}_{L}+D_{12}\Sigma^{0,1}_{L}\right) 
\left(\lambda_{1}^{R} D_{44}\Sigma^{1,0}_{R}+D_{43}\Sigma^{0,1}_{R}\right),
\end{align}
where 
\begin{align}
\lambda_{1}^{L}=\lambda_{1}^{R}=\left( -\frac{(\Delta_{\phi}-\Delta_{A}+\ell-1)(-\Delta_{\phi}+\Delta_{A}+d+\ell-1)}{(\Delta_{\phi}-\Delta_{A})(\Delta_{\phi}+\Delta_{A}-d)-(\ell-1)(d+\ell-1)} \right).
\end{align}
This leads to 
\begin{align}
g_{A}^{\Delta_{T\phi},\Delta_{\phi T},++++}(u,v)&=0,\\
g_{A}^{\Delta_{T\phi},\Delta_{\phi T},+3+3}(u,v)&=\frac12 [v\partial_{v}-\Delta_{T\phi}]g_{A}^{\Delta_{T\phi},\Delta_{\phi T},34}(u,v),\\
g_{A}^{\Delta_{T\phi},\Delta_{\phi T},34}(u,v)&=-\frac12 \sqrt{u}[1-\Delta_{T\phi}+v\partial_{v}]g_{\Delta_{A},\ell}^{\Delta_{T\phi}-1,\Delta_{\phi T}+1}(u,v)(1+O(1/\ell)),
\end{align}
where we used the approximation given in (\ref{eq:ablockdef}).
\subsection{Polarization Ratios}\label{sec:polratios}
Now we check that the different polarizations of the 4-point function $G(z,\bar{z})$ are related to each other by a $z$-independent factor. To see this we perform the sum over spins in the s-channel, via (\ref{eq:sumoverlint}). For $\langle J\phi\phi J\rangle$ this results in 
\begin{multline}
G_{J,STT}^{++}\propto -\frac{\Gamma(d)\Gamma(\Delta_{\phi})}{2^{4}}\sum_{n}(\lambda_{J\phi[J\phi]^{[\ell]}_{n}})^{2}F_{n}(u)\\-\frac{\Gamma(\frac{d}{2}+1)\Gamma(\Delta_{\phi}-\frac{d}{2}+1)}{2^{5}}(1-\bar{z})^{\frac{d}{2}-1}\sum_{n}(\lambda_{J\phi[J\phi]^{[\ell]}_{n}})^{2}\gamma_{[J\phi]_{n}^{[\ell]}}F_{n}(u)\ln(u),
\end{multline}
\begin{multline}
G_{J,STT}^{tt}\propto \frac{\Gamma(d-1)\Gamma(\Delta_{\phi})}{2^{5}}\sum_{n}(\lambda_{J\phi[J\phi]^{[\ell]}_{n}})^{2}F_{n}(u)\\+\frac{\Gamma(\frac{d}{2})\Gamma(\Delta_{\phi}-\frac{d}{2}+1)}{2^{6}}(1-\bar{z})^{\frac{d}{2}-1}\sum_{n}(\lambda_{J\phi[J\phi]^{[\ell]}_{n}})^{2}\gamma_{[J\phi]_{n}^{[\ell]}}F_{n}(u)\ln(u),
\end{multline}
where we defined 
\begin{align}
&F_{n}(u)\equiv \frac{2^{\Delta_{\phi}+d+2n}u^{n}}{\sqrt{\pi}(1-u)^{1-\frac{d}{2}}}{_{2}F_{1}}\left(\frac{d}{2}+n-1,\frac{d}{2}+n-1;\Delta_{\phi}+2n;u\right),\notag\\
&(\lambda_{J\phi[J\phi]^{[\ell]}_{n}})^{2}\equiv 2^{\ell}\ell^{-\frac12 (2\Delta_{\phi}+2d-7)}(\lambda_{J\phi[J\phi]^{[\ell]}_{n,\ell}})^{2},\quad \gamma_{[J\phi]_{n}^{[\ell]}}\equiv \ell^{d-2}\gamma_{[J\phi]_{n,\ell}^{[\ell]}},\notag
\end{align}
and used (\ref{eq:TwistJphiSTT}). The proportionality coefficient is the kinematical term in front of the 4-point function. The ratios $G^{tt}_{J,STT}/G^{++}_{J,STT}$ are then $\frac{1}{2(1-d)}$ at order $O((1-\bar{z})^{0})$ and $-\frac{1}{d}$ at order $O((1-\bar{z})^{\frac{d}{2}-1})$. Similarly, for $\langle T\phi\phi T \rangle$ we have 
\begin{multline}
G^{++++}_{T,STT}\propto\frac{\Gamma(d+2)\Gamma(\Delta_{\phi})}{2^{4}}\sum_{n}(\lambda_{T\phi[T\phi]^{[\ell]}_{n}})^{2}F_{n}(u)\\
+ \frac{\Gamma(\frac{d}{2}+3)\Gamma(\Delta_{\phi}-\frac{d}{2}+1)}{2^{5}}(1-\bar{z})^{\frac{d}{2}-1}\sum_{n}(\lambda_{T\phi[T\phi]^{[\ell]}_{n}})^{2}\gamma_{[T\phi]_{n}^{[\ell]}}F_{n}(u)\ln(u),
\end{multline}
\begin{multline}
G^{+3+3}_{T,STT}\propto-\frac{\Gamma(d+1)\Gamma(\Delta_{\phi})}{2^{5}}\sum_{n}(\lambda_{T\phi[T\phi]^{[\ell]}_{n}})^{2}F_{n}(u)\\
-\frac{\Gamma(\frac{d}{2}+2)\Gamma(\Delta_{\phi}-\frac{d}{2}+1)}{2^{6}}(1-\bar{z})^{\frac{d}{2}-1}\sum_{n}(\lambda_{T\phi[T\phi]^{[\ell]}_{n}})^{2}\gamma_{[T\phi]_{n}^{[\ell]}}F_{n}(u)\ln(u),
\end{multline}
\begin{multline}
G^{34}_{T,STT}\propto\frac{\Gamma(d)\Gamma(\Delta_{\phi})}{2^{5}}\sum_{n}(\lambda_{T\phi[T\phi]^{[\ell]}_{n}})^{2}F_{n}(u)\\
+\frac{\Gamma(\frac{d}{2}+1)\Gamma(\Delta_{\phi}-\frac{d}{2}+1)}{2^{6}}(1-\bar{z})^{\frac{d}{2}-1}\sum_{n}(\lambda_{T\phi[T\phi]^{[\ell]}_{n}})^{2}\gamma_{[T\phi]_{n}^{[\ell]}}F_{n}(u)\ln(u),
\end{multline}

\begin{multline}
G^{+3+3}_{T,A}\propto-\frac{\Gamma(d+1)\Gamma(\Delta_{\phi})}{2^{5}}\sum_{n}(\lambda_{T\phi[T\phi]^{[\ell,1]}_{n}})^{2}\tilde{F}_{n}(u)\\
-\frac{\Gamma(\frac{d}{2}+2)\Gamma(\Delta_{\phi}-\frac{d}{2}+1)}{2^{6}}(1-\bar{z})^{\frac{d}{2}-1}\sum_{n}(\lambda_{T\phi[T\phi]^{[\ell,1]}_{n}})^{2}\gamma_{[T\phi]_{n}^{[\ell,1]}}\tilde{F}_{n}(u)\ln(u),
\end{multline}
\begin{multline}
G^{34}_{T,A}\propto\frac{\Gamma(d)\Gamma(\Delta_{\phi})}{2^{4}}\sum_{n}(\lambda_{T\phi[T\phi]^{[\ell,1]}_{n}})^{2}\tilde{F}_{n}(u)\\
+\frac{\Gamma(\frac{d}{2}+1)\Gamma(\Delta_{\phi}-\frac{d}{2}+1)}{2^{5}}(1-\bar{z})^{\frac{d}{2}-1}\sum_{n}(\lambda_{T\phi[T\phi]^{[\ell,1]}_{n}})^{2}\gamma_{[T\phi]_{n}^{[\ell,1]}}\tilde{F}_{n}(u)\ln(u),
\end{multline}
where the twist for $A$ is given by (\ref{eq:TwistJphiA}) and 
\begin{align}
&\tilde{F}_{n}(u)\equiv \frac{2^{\Delta_{\phi}+d+2n}u^{n}}{\sqrt{\pi}(1-u)^{-\frac{d}{2}}}{_{2}F_{1}}\left(\frac{d}{2}+n,\frac{d}{2}+n;\Delta_{\phi}+2n+1;u\right),\notag\\
&(\lambda_{T\phi[T\phi]^{[\ell]}_{n}})^{2}\equiv 2^{\ell}\ell^{-\frac12 (2\Delta_{\phi}+2d-7)}(\lambda_{T\phi[T\phi]^{[\ell]}_{n,\ell}})^{2},\quad \gamma_{[T\phi]_{n}^{[\ell]}}\equiv \ell^{d-2}\gamma_{[T\phi]_{n,\ell}^{[\ell]}},\notag\\
&(\lambda_{T\phi[T\phi]^{[\ell,1]}_{n}})^{2}\equiv 2^{\ell-1}\ell^{-\frac12 (2\Delta_{\phi}+2d-5)}(\lambda_{T\phi[T\phi]^{[\ell,1]}_{n,\ell}})^{2},\quad \gamma_{[T\phi]_{n}^{[\ell,1]}}\equiv \ell^{d-2}\gamma_{[T\phi]_{n,\ell}^{[\ell,1]}}.\notag
\end{align}
For this case the ratios are summarized in table \ref{tab:ratios}.

\begin{table}[h]
	\caption{Ratios for the different polarizations of $\langle T\phi\phi T\rangle$  in the lightcone limit.}
	\label{tab:ratios}
	\centering

	\begin{tabular}{|c c c| }
	\hline
&$O((1-\bar{z})^{0})$&$O((1-\bar{z})^{\frac{d}{2}-1})$\\
	\hline
	$G^{+3+3}_{T,STT}/G^{++++}_{T,STT}$ & $-\frac{1}{2(d+1)}$&$-\frac{1}{d+4}$ \\
		$G^{34}_{T,STT}/G^{++++}_{T,STT}$ & $\frac{1}{2d(d+1)}$&$\frac{2}{(d+2)(d+4)}$ \\
$G^{34}_{T,A}/G^{+3+3}_{T,A}$ & $-\frac{2}{d}$&$-\frac{4}{d+2}$ \\
\hline
	\end{tabular}
\end{table}
\section{Anomalous Dimensions for Non-Zero $n$	}\label{sec:NonzeroN}
In this appendix we generalize our results for anomalous dimensions to $n > 0$ in the regime $\ell \gg n$. 
\subsection{$\langle J\phi\phi J \rangle $}
In order to match the identity at all orders in $z$, we use the summation formula 
 \begin{align}
 (1-x)^{b}=\sum_{n\geq 0}\frac{x^{n}(b)_{n}(c)_{n}}{n! (b+c+n-1)_{n}} {_{2}F_{1}}(b+n,b+n;b+c+2n;x)
 \end{align}
 in the s-channel expansion. This fixes the OPE coefficients, which we write in terms of the $n=0$ result:
\begin{align}
( \lambda_{J\phi[J\phi]^{[\ell]}_{n,\ell}} )^{2}=\frac{(1-\frac{d}{2}+\Delta_{\phi})_{n}(\frac{d}{2}-1)_{n}}{4^{n}n!(\Delta_{\phi}+n-1)_{n}}( \lambda_{J\phi[J\phi]^{[\ell]}_{0,\ell}} )^{2},
\end{align}
\begin{align}
( \lambda_{J\phi[J\phi]^{[\ell,1]}_{n,\ell}} )^{2}=\frac{(1-\frac{d}{2}+\Delta_{\phi})_{n}(\frac{d}{2})_{n}}{4^{n}n!(\Delta_{\phi}+n)_{n}}( \lambda_{J\phi[J\phi]^{[\ell,1]}_{0,\ell}} )^{2}.
\end{align}

Now we split the anomalous dimensions as 
\begin{align}
\gamma_{[J\phi]_{n,\ell}^{[\ell]}}=\tilde{\gamma}_{[J\phi]_{n}^{[\ell]}}\gamma_{[J\phi]_{0,\ell}^{[\ell]}},\quad \gamma_{[J\phi]_{n,\ell}^{[\ell,1]}}=\tilde{\gamma}_{[J\phi]_{n}^{[\ell,1]}}\gamma_{[J\phi]_{0,\ell}^{[\ell,1]}},
\end{align}
and match the stress-tensor at all orders in $z$. This leads to the following equations 
\begin{multline}\label{eq:AnomDimEq1}
\frac{\Gamma(\frac{d}{2}-1)\Gamma(\Delta_{\phi}-\frac{d}{2}+1)(\frac{d}{2}+1)_{j}^{2}}{(j!)^{2}\Gamma(\Delta_{\phi}-\frac{d}{2}+1+j)^{2}}{_{3}F_{2}}\left(\begin{matrix}
	-j,-j,\Delta_{\phi}-d\\ 
	-\frac{d}{2}-j,-\frac{d}{2}-j
\end{matrix};1\right)\\
=\sum_{n=0}^{j}\frac{(\Delta_{\phi}+2n-1)\Gamma(\frac{d}{2}+n-1)\Gamma(\Delta_{\phi}+n-1)}{n!(j-n)\Gamma(\Delta_{\phi}+n-\frac{d}{2}+1)\Gamma(\Delta_{\phi}+n+j)}\tilde{\gamma}_{[J\phi]_{n}^{[\ell]}},
\end{multline}
\begin{multline}\label{eq:AnomDimEq2}
\frac{\Gamma(\frac{d}{2})\Gamma(\Delta_{\phi}-\frac{d}{2}+1)(\frac{d}{2}+1)_{j}^{2}}{(j!)^{2}\Gamma(\Delta_{\phi}-\frac{d}{2}+1+j)^{2}}{_{3}F_{2}}\left(\begin{matrix}
	-j,-j,\Delta_{\phi}-d\\ 
	-\frac{d}{2}-j,-\frac{d}{2}-j
\end{matrix};1\right)\\
=\sum_{n=0}^{j}\frac{(\Delta_{\phi}+2n)\Gamma(\frac{d}{2}+n)\Gamma(\Delta_{\phi}+n)}{n!(j-n)\Gamma(\Delta_{\phi}+n-\frac{d}{2}+1)\Gamma(\Delta_{\phi}+n+j+1)}\tilde{\gamma}_{[J\phi]_{n}^{[\ell,1]}},
\end{multline}
where $j$ represents the power of $z$ in the Taylor expansion.
Using the techniques of \cite{Kaviraj:2015xsa,Kaviraj:2015cxa}, we write $\tilde{\gamma}$ in terms of terminating hypergeometric functions:
\begin{multline}
\tilde{\gamma}_{[J\phi]_{n}^{[\ell]}}=\frac{(-1)^{n}n!\Gamma(\Delta_{\phi}-\frac{d}{2}+1)\Gamma(\Delta_{\phi}+n-\frac{d}{2}+1)}{(\frac{d}{2}-1)_{n}\Gamma(\frac{d}{2}+1)^{2}}\\\times \sum_{i=0}^{n}
\frac{(-1)^{i}(i+1)_{\frac{d}{2}}^{2}(\Delta_{\phi}+n-1)_{i}}{(n-i)!\Gamma(\Delta_{\phi}-\frac{d}{2}+1+i)^{2}}
{_{3}F_{2}}\left(\begin{matrix}
	-i,-i,\Delta_{\phi}-d\\ 
	-\frac{d}{2}-i,-\frac{d}{2}-i
\end{matrix};1\right),
\end{multline}
\begin{multline}
\tilde{\gamma}_{[J\phi]_{n}^{[\ell,1]}}=\frac{(-1)^{n}n!\Gamma(\Delta_{\phi}-\frac{d}{2}+1)\Gamma(\Delta_{\phi}+n-\frac{d}{2}+1)}{(\frac{d}{2})_{n}\Gamma(\frac{d}{2}+1)^{2}}\\\times \sum_{i=0}^{n}
\frac{(-1)^{i}(i+1)_{\frac{d}{2}}^{2}(\Delta_{\phi}+n)_{i}}{(n-i)!\Gamma(\Delta_{\phi}-\frac{d}{2}+1+i)^{2}}
{_{3}F_{2}}\left(\begin{matrix}
	-i,-i,\Delta_{\phi}-d\\ 
	-\frac{d}{2}-i,-\frac{d}{2}-i
\end{matrix};1\right).
\end{multline}
One can check that this solves (\ref{eq:AnomDimEq1}) and (\ref{eq:AnomDimEq2}) order by order in $n$, for arbitrarily high values.
\subsection{$\langle T\phi\phi T \rangle $}
Following the same steps as in the previous case, we find the OPE coefficients
\begin{align}
( \lambda_{T\phi[T\phi]^{[\ell]}_{n,\ell}} )^{2}=\frac{(1-\frac{d}{2}+\Delta_{\phi})_{n}(\frac{d}{2}-1)_{n}}{4^{n}n!(\Delta_{\phi}+n-1)_{n}}( \lambda_{T\phi[T\phi]^{[\ell]}_{0,\ell}} )^{2},
\end{align}
\begin{align}
( \lambda_{T\phi[T\phi]^{[\ell,1]}_{n,\ell}} )^{2}=\frac{(1-\frac{d}{2}+\Delta_{\phi})_{n}(\frac{d}{2})_{n}}{4^{n}n!(\Delta_{\phi}+n)_{n}}( \lambda_{T\phi[T\phi]^{[\ell,1]}_{0,\ell}} )^{2},
\end{align}
\begin{align}
( \lambda_{T\phi[T\phi]^{[\ell,2]}_{n,\ell}} )^{2}=\frac{(1-\frac{d}{2}+\Delta_{\phi})_{n}(\frac{d}{2}+1)_{n}}{4^{n}n!(\Delta_{\phi}+n+1)_{n}}( \lambda_{T\phi[T\phi]^{[\ell,2]}_{0,\ell}} )^{2}.
\end{align}
Notice that for $[\ell]$ and $[\ell,1]$, the n-dependence is the same as in $\langle J\phi\phi J\rangle$.
Finally, we define anomalous dimensions for $n\geq 0$ as
\begin{align}
\gamma_{[T\phi]_{n,\ell}^{[\ell]}}=\tilde{\gamma}_{[T\phi]_{n}^{[\ell]}}\gamma_{[T\phi]_{0,\ell}^{[\ell]}},\quad \gamma_{[T\phi]_{n,\ell}^{[\ell,1]}}=\tilde{\gamma}_{[T\phi]_{n}^{[\ell,1]}}\gamma_{[T\phi]_{0,\ell}^{[\ell,1]}}, \quad \gamma_{[T\phi]_{n,\ell}^{[\ell,2]}}=\tilde{\gamma}_{[T\phi]_{n}^{[\ell,2]}}\gamma_{[T\phi]_{0,\ell}^{[\ell,2]}}.
\end{align}
For $STT$ and $A$ we find the same equations as in $\langle J\phi\phi J\rangle$. Therefore $\tilde{\gamma}_{[T\phi]_{n}^{[\ell]}}=\tilde{\gamma}_{[J\phi]_{n}^{[\ell]}}$ and $\tilde{\gamma}_{[T\phi]_{n}^{[\ell,1]}}=\tilde{\gamma}_{[J\phi]_{n}^{[\ell,1]}}$. On the other hand, for $B$ we have 
\begin{multline}
\frac{\Gamma(\frac{d}{2}+1)\Gamma(\Delta_{\phi}-\frac{d}{2}+1)(\frac{d}{2}+1)_{j}^{2}}{(j!)^{2}\Gamma(\Delta_{\phi}-\frac{d}{2}+1+j)^{2}}{_{3}F_{2}}\left(\begin{matrix}
	-j,-j,\Delta_{\phi}-d\\ 
	-\frac{d}{2}-j,-\frac{d}{2}-j
\end{matrix};1\right)\\
=\sum_{n=0}^{j}\frac{(\Delta_{\phi}+2n+1)\Gamma(\frac{d}{2}+n+1)\Gamma(\Delta_{\phi}+n+1)}{n!(j-n)\Gamma(\Delta_{\phi}+n-\frac{d}{2}+1)\Gamma(\Delta_{\phi}+n+j+2)}\tilde{\gamma}_{[T\phi]_{n}^{[\ell,2]}}.
\end{multline}
The solution is 
\begin{multline}
\tilde{\gamma}_{[T\phi]_{n}^{[\ell,2]}}=\frac{(-1)^{n}n!\Gamma(\Delta_{\phi}-\frac{d}{2}+1)\Gamma(\Delta_{\phi}+n-\frac{d}{2}+1)}{(\frac{d}{2}+1)_{n}\Gamma(\frac{d}{2}+1)^{2}}\\\times \sum_{i=0}^{n}
\frac{(-1)^{i}(i+1)_{\frac{d}{2}}^{2}(\Delta_{\phi}+n+1)_{i}}{(n-i)!\Gamma(\Delta_{\phi}-\frac{d}{2}+1+i)^{2}}
{_{3}F_{2}}\left(\begin{matrix}
	-i,-i,\Delta_{\phi}-d\\ 
	-\frac{d}{2}-i,-\frac{d}{2}-i
\end{matrix};1\right).
\end{multline}
\subsection{Examples}
Now using the identities in the appendices of \cite{Kaviraj:2015xsa,Kaviraj:2015cxa} we can rewrite the terminating hypergeometric and perform the sum over $i$ for specific even dimensions. In $d=4$ we have 
\begin{align}
\tilde{\gamma}_{[T\phi]_{n}^{[\ell]}}&=1+\frac{3n(\Delta_{\phi}+n-1)(\Delta_{\phi}+n(\Delta_{\phi}+n-1))}{\Delta_{\phi}(\Delta_{\phi}-1)},\notag\\
\tilde{\gamma}_{[T\phi]_{n}^{[\ell,1]}}&=\frac{(n+1)(\Delta_{\phi}+n-1)(\Delta_{\phi}+n(\Delta_{\phi}+n))}{\Delta_{\phi}(\Delta_{\phi}-1)},\\
\tilde{\gamma}_{[T\phi]_{n}^{[\ell,2]}}&=\frac{(n+1)(n+2)(\Delta_{\phi}+n-1)(\Delta_{\phi}+n)}{2\Delta_{\phi}(\Delta_{\phi}-1)},\notag
\end{align}
whereas in $d=6$
\begin{align}
\tilde{\gamma}_{[T\phi]_{n}^{[\ell]}}&=\frac{(n+1)(\Delta_{\phi}+n-2)(5n^{2}(n-1)^{2}+2\Delta_{\phi}(5n^{3}-5n-3)+\Delta_{\phi}^{2}(5n(n+2)+6))}{6\Delta_{\phi}(\Delta_{\phi}-1)(\Delta_{\phi}-2)},\notag\\
\tilde{\gamma}_{[T\phi]_{n}^{[\ell,1]}}&=\frac{(n+1)(n+2)(\Delta_{\phi}+n-2)(\Delta_{\phi}+n-1)(3\Delta_{\phi}+2n(\Delta_{\phi}+n))}{6\Delta_{\phi}(\Delta_{\phi}-1)(\Delta_{\phi}-2)},\\
\tilde{\gamma}_{[T\phi]_{n}^{[\ell,2]}}&=\frac{(n+1)(n+2)(n+3)(\Delta_{\phi}+n-2)(\Delta_{\phi}+n-1)(\Delta_{\phi}+n)}{6\Delta_{\phi}(\Delta_{\phi}-1)(\Delta_{\phi}-2)}.\notag
\end{align}
It is easy to check that for $\Delta_{\phi}\geq \frac{d}{2}-1$ these expressions are positive for all $n$.
\section{Correlation Functions of Conserved Operators}
\label{sec:TensorBasis}
In this appendix we will provide more details on the 3-point functions $\<JJJ\>$ and $\<TTT\>$.
\subsection{Tensor Structures}
We follow the notation and techniques of \cite{Costa:2011dw,Costa:2011mg}. See also \cite{Li:2015itl} for more details on the differential representation on the 3-point functions. 

We start by lifting the points $x^{\mu}$ and polarizations $\epsilon^{\mu}$ in $R^{d-1,1}$ to vectors $P^{\mu}$ and $Z^{\mu}$, respectively, in $R^{d,2}$.
This lifting allows us to linearize the action of the conformal group. The projection from embedding space to the Poincar\'{e} section is explicitly given by,
\begin{align}
P_{x}=(P^{+},P^{-},P_{\mu})=(1,x^{2},x^{\mu}), \qquad Z_{\epsilon}=(0,2x\cdot\epsilon,\epsilon^{\mu})
\end{align}
The metric on the embedding space is given by $P\cdot P=-P^{+}P^{-}+\delta_{ab}P^{a}P^{b}$. 
The basic buildings blocks are given by
\begin{align}
H_{ij}=-2[(Z_{i}\cdot Z_{j})(P_{i}\cdot P_{j}-(Z_{i}\cdot P_{j})(Z_{j}\cdot P_{i}))], \quad
V_{i,jk}=\frac{(Z_{i}\cdot P_{j})(P_{i}\cdot P_{k})-(Z_{i}\cdot P_{k})(P_{i}\cdot P_{j})}{P_{j}\cdot P_{k}}.
\end{align}
We will use the shorthand $V_{1}=V_{1,23}$, $V_{2}=V_{2,31}$, and $V_{3}=V_{3,1,2}$.

\subsection{$\<JJT\>$}
We will normalize the operators as follows:
\begin{align}
\<J(P_{1},Z_{1})J(P_{2},Z_{2})\>=C_{J}\frac{H_{12}}{P_{12}^{d}}, \quad \<T(P_{1},Z_{1})T(P_{2},Z_{2})\>=C_{T}\frac{H_{12}^{2}}{P_{12}^{d+2}} .
\end{align}

The general form of the $\<JJT\>$ 3-point function, after imposing symmetry under $1\leftrightarrow 2$, is given by
\bea
\<J(P_{1};Z_{1})J(P_{2};Z_{2})T(P_{3};Z_{3})\>=\frac{\alpha V_{1}V_{2}V_{3}^{2}+\beta(H_{13}V_{2}+H_{23}V_{1})V_{3}+\gamma H_{12}V_{3}^{2}+\eta H_{13}H_{23}}{(P_{12})^{\frac{d}{2}-1}(P_{13})^{\frac{d}{2}+1}(P_{23})^{\frac{d}{2}+1}} .\nonumber
\eea
Imposing conservation implies
\bea
-\alpha-d\beta+(2+d)\gamma=0, \nonumber \\
-2\beta+2\gamma+(2-d)\eta=0.
\eea

The relation between our basis and that used in \cite{Osborn:1993cr}, see Eqs.~(3.11-3.14) is given by\footnote{We add tildes to the variables to avoid confusion between these variables, the conformal anomalies $a$ and $c$, and the $\<TTT\>$ OPE coefficients.}
\bea
\begin{split}
&\eta=2\tilde{e}, \ \ \ \beta=-2\tilde{c},  \\
&\gamma= \tilde{a}-\frac{\tilde{b}}{d}-\frac{4\tilde{c}}{d}, \ \ \   \alpha=2\tilde{a}+\tilde{b}\big(1-\frac{2}{d}\big)-\frac{8\tilde{c}}{d},
\end{split}
\eea
They also found that the Ward identity for the stress energy tensor implies
\bea
2S_{d}(\tilde{c}+\tilde{e})=dC_{J}.
\eea
\noindent Where $S_{d}$ is the volume of a $(d-1)$-dimensional sphere, $S_{d}=\frac{2\pi^{\frac{d}{2}}}{\Gamma(\frac{d}{2})}$. So $\<JJT\>$ is fixed up to one OPE coefficient, $\tilde{c}$, and $C_{J}$. We labeled the parameter $\tilde{c}$ as $\lambda_{JJT}$ in the body of the paper, following the conventions of \cite{Li:2015itl}. In the rest of this appendix we will also adopt this convention. To construct the conformal block corresponding to $T^{\mu\nu}$ exchange in the s-channel of $\<JJ\f\f\>$ we apply the following differential operator on the scalar partial wave,
\bea
D_{L,T}
&=\bigg[\bigg(2\lambda_{JJT}-\frac{C_{J}d(d-2)}{(d-1)S_{d}}\bigg)D_{11}D_{22}+\bigg(2\lambda_{JJT}+\frac{C_{J}d^{2}}{S_{d}(1-d)}\bigg)D_{12}D_{21}-2\lambda_{JJT}H_{12}\bigg]\Sigma_{L}^{1,1}. \hspace{0.5cm}
\eea
The conformal block for $T^{\mu\nu}$ exchange in the t-channel of $\<J\f\f J\>$ is then found by letting $2\leftrightarrow 4$ everywhere in the resulting expression. 

Finally, in \cite{Hofman:2008ar} the parameter $a_{2}$ was introduced, distinct from the $a_{2}$ OPE coefficient used in $\<VVT\>$, which gives the energy distribution for a state created by a conserved current: 
\bea
\<\mathcal{E}(n)\>_{\epsilon\cdot j}=\frac{1}{S_{d}}\bigg(1+a_{2}\big(\text{cos}^{2}(\theta)-\frac{1}{d-1}\big)\bigg)
\eea
where $\theta$ is the angle between the spatial polarization $\epsilon^{i}$ and the point on $S^{d-1}$ labelled by $n^{i}$. Requiring that the energy one point function be positive yields the bounds
\bea
-\frac{d-1}{d-2}\leq a_{2}\leq d-1 .
\eea 
The upper bound is saturated in a theory of free bosons and the lower bound is saturated in a theory of free fermions. The relation between $\lambda_{JJT}$ and $a_{2}$ is given by
\bea
\lambda_{JJT}=-\frac{C_{J}(d-2) d \pi ^{-\frac{d}{2}} \left(a_{2}-d^2+d\right) \Gamma \left(\frac{d}{2}\right)}{4 (d-1)^3} .
\eea

\subsection{$\<TTT\>$}
\label{sect:TTT}
In this section we will review the connection between the parametrization of $\<TTT\>$ in terms of the variables $\hat{c}$, $\hat{e}$, and $C_{T}$ as defined in \cite{Osborn:1993cr}, the $t_{2}$, $t_{4}$, $C_{T}$ parametrization used in studies of the energy one point function \cite{Buchel:2009sk}, and the free field theory results. 

We start by defining the following basis of parity-even tensor structures for $\<TTT\>$,
\begin{align}
&Q_{1}=V_{1}^{2}V_{2}^{2}V_{3}^{2}, \label{eqn:stand1} \\
&Q_{2}=H_{23}V_{1}^{2}V_{2}V_{3}+H_{13}V_{1}V_{2}^{2}V_{3},\\
&Q_{3}=H_{12}V_{1}V_{2}V_{3}^{2},\\
&Q_{4}=H_{12}H_{13}V_{2}V_{3}+H_{12}H_{23}V_{1}V_{3},\\
&Q_{5}=H_{13}H_{23}V_{1}V_{2},\\
&Q_{6}=H_{12}^{2}V_{3}^{2},\\
&Q_{7}=H_{13}^{2}V_{2}^{2}+H_{23}^{2}V_{1}^{2},\\
&Q_{8}=H_{12}H_{13}H_{23}. \label{eqn:stand8}
\end{align}

In \cite{Osborn:1993cr} they parametrized the correlation function in general dimensions in terms of 8 variables: $\hat{a}$, $\hat{b}$, $\hat{b}'$, $\hat{c}$, $\hat{c}'$, $\hat{e}$, $\hat{e}'$, and $\hat{f}$. Labeling the coefficients of $Q_{i}$ by $x_{i}$, the relation between the bases is given by
\bea
x_{1}=8(\hat{c}+\hat{e})+\hat{f}, \ \ \
x_{2}=-4(4\hat{b}'+\hat{e}'),\ \ \
x_{3}=4(2\hat{c}+\hat{e}),\\
x_{4}=-8\hat{b}', \qquad \qquad
x_{5}=8\hat{b}+16\hat{a}, \qquad \qquad \ \ 
x_{6}=2\hat{c},\\
x_{7}=2\hat{c}', \qquad \qquad \qquad
x_{8}=8\hat{a}.
\eea

Conservation of the stress-energy tensor implies
\bea
x_{1}=2x_{2}+\frac{1}{4}(d^{2}+2d-8)x_{4}-\frac{1}{2}d(2+d)x_{7}, \qquad
x_{8}=\frac{x_{2}-(\frac{d}{2}+1)x_{4}+2dx_{7}}{\frac{d^{2}}{2}-2},\\
x_{2}=x_{3},\qquad 
x_{4}=x_{5},\qquad 
x_{6}=x_{7},
\eea
which is consistent with the conservation constraints of \cite{Osborn:1993cr}. Finally, they found that solving the Ward identity yields 
\bea
4S_{d}\frac{(d-2)(d+3)\hat{a}-2\hat{b}-(d+1)\hat{c}}{d(d+2)}=C_{T}.
\eea
In $d>3$ dimensions we can parametrize the parity-even structures in $\<TTT\>$ by $\hat{c}$, $\hat{e}$, and $C_{T}$, while in $d=3$ the $H_{12}H_{13}H_{23}$ structure is not linearly independent and $\<TTT\>$ is fixed up to two parameters, $2\hat{a}-\hat{c}$ and $C_{T}$. 

The relation between this basis and the $t_{2}$ and $t_{4}$ basis is given by
\begin{align}
\hat{c}&=-\frac{C_{T} \pi ^{-\frac{d}{2}} \Gamma \left(\frac{d}{2}+2\right) \left[\left(d \left(-3 d^2+d+2\right)+4\right) t_{4}+(d+1) \left(2 d^4-d^3 (t_{2}+4)+d^2+d+3 t_{2}\right)\right]}{2 (d-1)^3 (d+1)^2 (d+2)},
\\
\hat{e}&=\frac{C_{T} \pi ^{-\frac{d}{2}} \Gamma \left(\frac{d}{2}+2\right) \left[(d+1) \left((d-3) \left(d^2-3\right) t_{2}+2 (d-2) d^2+2\right)+(2(d-5) d^2+4d+12) t_{4}\right]}{4 (d-1)^3 (d+1)^2}.
\end{align}

Finally, as noted in \cite{Buchel:2009sk}, in even dimensions we can parametrize $\<TTT\>$ by its expressions in free field theories of conformally coupled scalars, fermions, and $(\frac{d}{2}-1)$-forms:
\bea
\<TTT\>=n_{s}\<TTT\>_{s}+n_{f}\<TTT\>_{f}+n_{t}\<TTT\>_{t},
\eea
where $n_{s}$, $n_{f}$, and $n_{t}$ give the effective number of real scalars, Dirac fermions, and $(\frac{d}{2}-1)$ forms, although there may not necessarily be any connection to the actual field content. 
The conformal collider constraints can then be written as \cite{Camanho:2009vw,Buchel:2009sk},
\begin{align}
&\bigg(1-\frac{1}{d-1}t_{2}-\frac{2}{d^{2}-1}t_{4}\bigg)+\frac{d-2}{d-1}(t_{2}+t_{4})\propto n_{s} \geq0, \label{eq:TTTCollider1}
\\
&\bigg(1-\frac{1}{d-1}t_{2}-\frac{2}{d^{2}-1}t_{4}\bigg)+\frac{1}{2}t_{2}\propto n_{f}\geq0, \label{eq:TTTCollider2}
\\
&\bigg(1-\frac{1}{d-1}t_{2}-\frac{2}{d^{2}-1}t_{4}\bigg)\propto n_{t}\geq0. \label{eq:TTTCollider3}
\end{align}
The constraints (\ref{eq:TTTCollider1}), (\ref{eq:TTTCollider2}), and (\ref{eq:TTTCollider3}) are equivalent to the constraints derived by considering $\<T^{++}\f\f T^{++}\>$, $\<T^{+3}\f\f T^{+3}\>$, and $\<(T^{33}-T^{44})\f\f (T^{33}-T^{44})\>$, respectively. In three dimensions $t_{2}=0$ and the second and third constraints are redundant.
Finally in four dimensions we have  \cite{Duff:1977ay,Hofman:2008ar,Osborn:1993cr}
\bea
\frac{a}{c}=\frac{2n_{s}+124n_{t}+22n_{f}}{6n_{s}+72n_{t}+36n_{f}},
\eea
where $n_{t}$ now counts the number of real free vectors, $a$ is the Euler anomaly, and $c$ is related to central charge $C_{T}$ as $c=\frac{\pi^{4}}{40}C_{T}$. The bounds from equations (\ref{eq:TTTCollider1}) and (\ref{eq:TTTCollider3}) then imply 
\bea
\frac{31}{18}\geq\frac{a}{c}\geq\frac{1}{3}.
\eea

\bibliography{Biblio}{}
\bibliographystyle{utphys}

\end{document}